\documentclass[12pt,a4paper]{article}
\usepackage{jheppub}
\usepackage{dsfont}
\usepackage{amssymb,amsmath}
\usepackage{epsfig}
\usepackage{epstopdf}
\usepackage{latexsym}
\usepackage{graphicx}
\usepackage{subfigure}
\usepackage{booktabs}
\usepackage{bbm}
\pdfoutput=1
\makeatletter
\def\@fpheader{\relax}
\makeatother

\def\F{{\cal F}}
\def\cM{{\cal M}}
\def\cN{{\cal N}}

\def\S{{\cal S}}
\def\A{{\cal A}}
\def\R{{\cal R}}
\def\L{{\cal L}}
\def\cO{{\cal O}}

\def\F{{\cal F}}
\def\C{{\cal C}}
\def\X{{\cal X}}
\def\Z{{\cal Z}}
\def\D{{\cal D}}

\title{Maximal Chaos from Strings, Branes and Schwarzian Action}
\author{Avik Banerjee$^{a}$, Arnab Kundu$^{a}$, Rohan Poojary$^b$}
\affiliation{$^a$Theory Division, Saha Institute of Nuclear Physics, HBNI, 1/AF Bidhannagar, Kolkata 700064, India.}
\affiliation{$^b$Department of Theoretical Physics, Tata Institute of Fundamental Research, Mumbai 400005, India.}
\emailAdd{avik.banerjee [at] saha.ac.in} 
\emailAdd{arnab.kundu [at] saha.ac.in} 
\emailAdd{ronp [at] theory.tifr.res.in}

\abstract{In this article, we explicitly demonstrate that, for a sufficiently generic class of examples, an open string embedded in an AdS-background yields an effective Schwarzian action and, in the semi-classical description, the fluctuation modes of the open string couple to this Schwarzian sector. This leads to a maximal chaos, observed in the open string degrees of freedom, irrespective of the gravitational background. This corresponds to the dynamics of quark-like degrees of freedom in a strongly coupled large $N_c$ gauge theory. We also demonstrate a maximal chaos, resulting from an inherent D-brane horizon, by computing the four-point out-of-time-ordered correlator of spin-one operators. We also offer some observations and comments regarding a class of effective theories that is described by a generic functional of a Schwarzian derivative.}

\begin{document}

\maketitle
\flushbottom

\section{Introduction}

We will begin reviewing some basic ingredients that we will need for the rest of the article. The canonical statement of holographic duality, specially in the context of conformal field theories with large central charge, schematically takes the following form:
\begin{eqnarray}
\Z_{\rm gravity} = \Z_{\rm QFT} \ , \label{adscft}
\end{eqnarray}
where ``gravity" implies a classical diffeomorphism invariant theory of fields, ``QFT" implies a general quantum field theory with large number of degrees of freedom and $\Z$ generically represents the path integral in Lorentzian or Euclidean signature. The conjecture extends naturally to complex valued time-direction as well, which, in general, corresponds to Schwinger-Keldysh contours in QFT. Correspondingly, one can define a bulk realization of a given Schwinger-Keldysh contour.

Given the definition in (\ref{adscft}), one calculates {\it e.g.}~the Euclidean correlation functions {\it via} the GKPW prescription\cite{Gubser:1998bc, Witten:1998qj}. Towards that, the gravity Euclidean path is schematically defined as:
\begin{eqnarray}
\Z_{\rm gravity} \left[G_0, \phi_0 \right] = \int \D G \D \phi \, e^{- S_{\rm E} \left[ G, \phi \right ]} \ ,
\end{eqnarray}
where $\{G, \phi\}$ are collective bulk gravitational data, $\{G_0, \phi_0\}$ are the corresponding boundary data at the conformal boundary of AdS and $S_{\rm E}$ is the Euclidean action. The Euclidean path integral is defined over a particular integration measure, which we do not explicitly consider. The duality in (\ref{adscft}) then ensures:
\begin{eqnarray}
\Z_{\rm gravity} \left[G_0, \phi_0 \right] \equiv \langle e^{- \int \phi_0 \cO + \ldots}\rangle_{\rm QFT} \ ,
\end{eqnarray}
where $\cO$ is the operator corresponding to the source $\phi_0$. The above equivalence, clearly, holds at a saddle point of the corresponding action. Thus, an $n$-point correlation function is simply computed by considering various fluctuation fields around this classical saddle, using:
\begin{eqnarray}
\langle \cO_1 (x_1) \ldots \cO_n (x_n)\rangle = \left. (-1)^{n+1} \frac{\delta^n S_{\rm E}^{\rm on-shell}} {\delta \phi_0^1 (x_1) \ldots \delta \phi_0^n (x_n)} \right |_{\phi_0^i = 0 } \ ,
\end{eqnarray}
where the operator $\cO_i$ is sourced by the field $\phi_0^i$ at the conformal boundary.

\begin{figure}[ht!]
\begin{center}
{\includegraphics[width=0.8\textwidth]{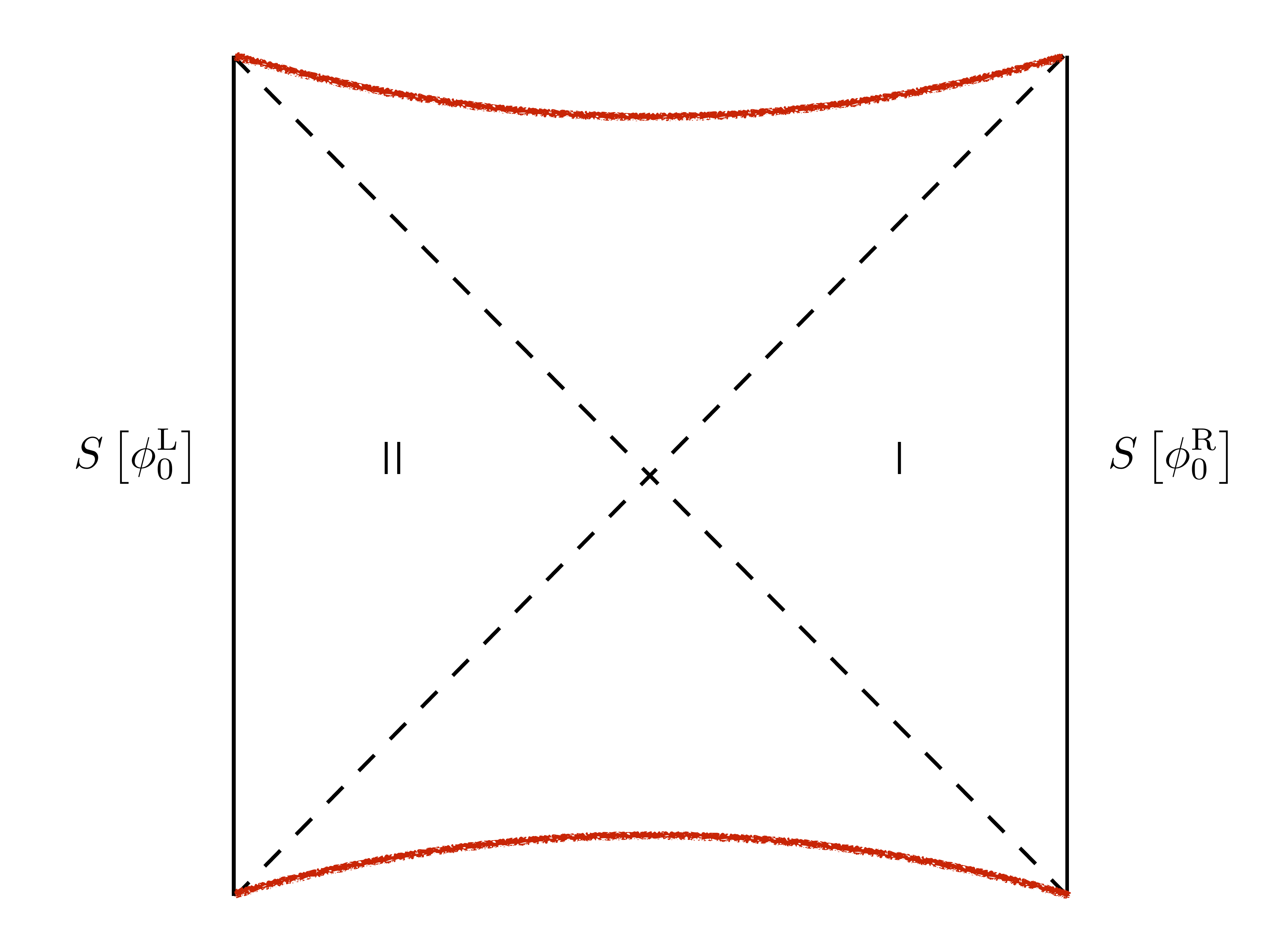}}
\caption{\small The maximally extended geometry for an eternal black hole in AdS. This represents a pure state in which the complete dual CFT is a tensor product of the two CFTs, defined on the left and the right boundaries.} \label{kruskal}
\end{center}
\end{figure}

In the Lorentzian picture, the corresponding $n$-point correlation function can also be calculated using a similar prescription. However, the Lorenztian patch comes equipped with an additional feature. Suppose, we consider a saddle which is described by a black hole in the Lorentzian section. In the saddle, the maximal analytic extension of the black hole geometry, which is provided by the Kruskal extension, contains two conformal boundaries. This is a close analogue of the Schwinger-Keldysh picture in quantum field theory and was explored first in \cite{Herzog:2002pc}, in the context of AdS/CFT correspondence. Pictorially, this is represented in figure \ref{kruskal}.
Thus, given an Euclidean section, the corresponding thermal state can be thought of as obtained from a two-sided pure state (in the corresponding Lorentzian patch). The corresponding thermofield double (TFD) state is given by
\begin{eqnarray}
\left | {\rm TFD} \right \rangle = \frac{1}{\Z[\beta]} \sum_n e^{- \beta E_n /2} \left | n \right \rangle_{\rm L} \left | n \right \rangle_{\rm R} \ ,
\end{eqnarray}
where $\left | n \right \rangle_{\rm L}$ and $\left | n \right \rangle_{\rm R}$ are energy eigenstates of the left and the right QFTs, whose actions are denoted by $S\left[ \phi_0^{\rm L}\right]$ and $S\left[ \phi_0^{\rm R}\right]$ in figure \ref{kruskal}, respectively. The thermal density matrix of {\it e.g.}~the right QFT is then obtained from tracing out the ``left" QFT degrees of freedom in the pure density matrix $\rho_{\rm TFD} = \left | {\rm TFD} \right \rangle \left \langle {\rm TFD} \right |$. Running this argument backwards, given a thermal density matrix one can double the degrees of freedom such that the extended density matrix becomes a pure one.

On the other hand, an Euclidean path integral, can be interpreted as a state, denoted by $\left | \psi \right \rangle$ in the Schr\"{o}dinger picture. Such a state can then be time-evolved according to an unitary operator $U\left( t, t_0\right) $.
\begin{figure}[ht!]
\begin{center}
{\includegraphics[width=0.8\textwidth]{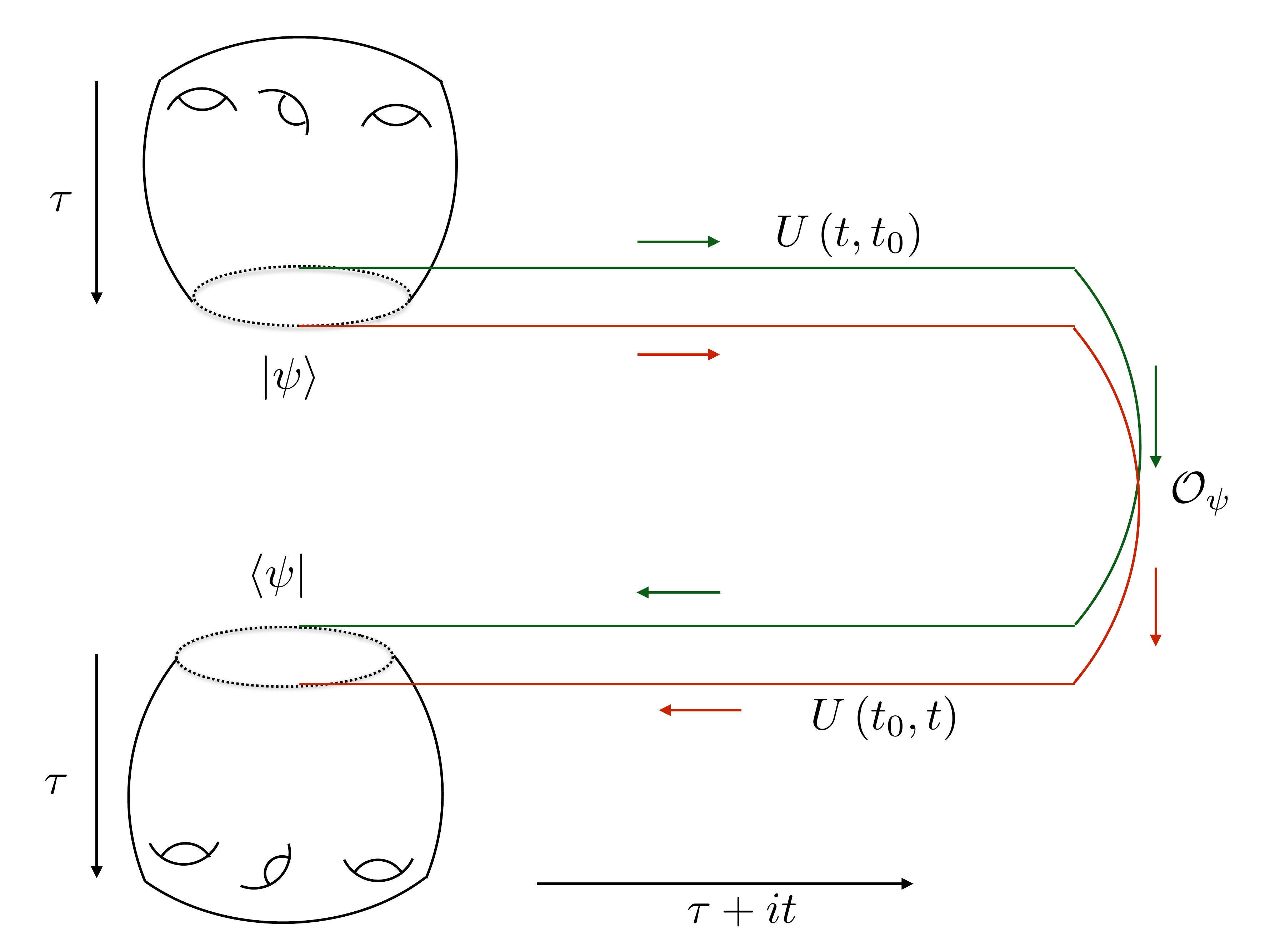}}
\caption{\small A pictorial representation of gluing two arbitrary states along a Schwinger-Keldysh contour, after the insertion of an operator at real time $t$. The genus $3$ surface in the picture is supposed to represent an arbitrary initial state. The choice of genus $=3$ is simply representative. This picture is inspired from the discussions and figures in \cite{SonnerLecture}.} \label{SK}
\end{center}
\end{figure}
The expectation value of an one-point function is thus given by
\begin{eqnarray}
\langle \cO_\psi (t ) \rangle =  \langle \psi(t)  | \cO_\psi | \psi(t) \rangle = \langle \psi  | U\left( t, t_0 \right) \cO_\psi U\left( t_0, t \right) | \psi \rangle\ .
\end{eqnarray}
The second line above is pictorially presented in figure \ref{SK}. For a more detailed discussion, see {\it e.g.}~\cite{SonnerLecture}.

One can easily generalize this prescription for higher point functions. For example, consider a general four-point function, without any time-ordering\cite{Shenker:2013pqa, Shenker:2013yza, Shenker:2014cwa, Maldacena:2015waa}:
\begin{eqnarray}
\C_{(4)} = \langle V(0) W(t) V(0) W(t) \rangle = \langle 0 | e^{-\beta H} V(0) e^{-iHt} W(0) e^{iHt} V(0) e^{- iHt} W(0) e^{iHt} e^{-\beta H} |0 \rangle \ . \nonumber\\
\end{eqnarray}
This require us to first prepare the thermal state by $e^{-\beta H} | 0 \rangle$, and subsequently choosing a time contour according to the Hamiltonian evolution inside the correlation function, from right to left; with the insertion of the corresponding operator $V$ and $W$.   
\begin{figure}[ht!]
\begin{center}
{\includegraphics[width=0.8\textwidth]{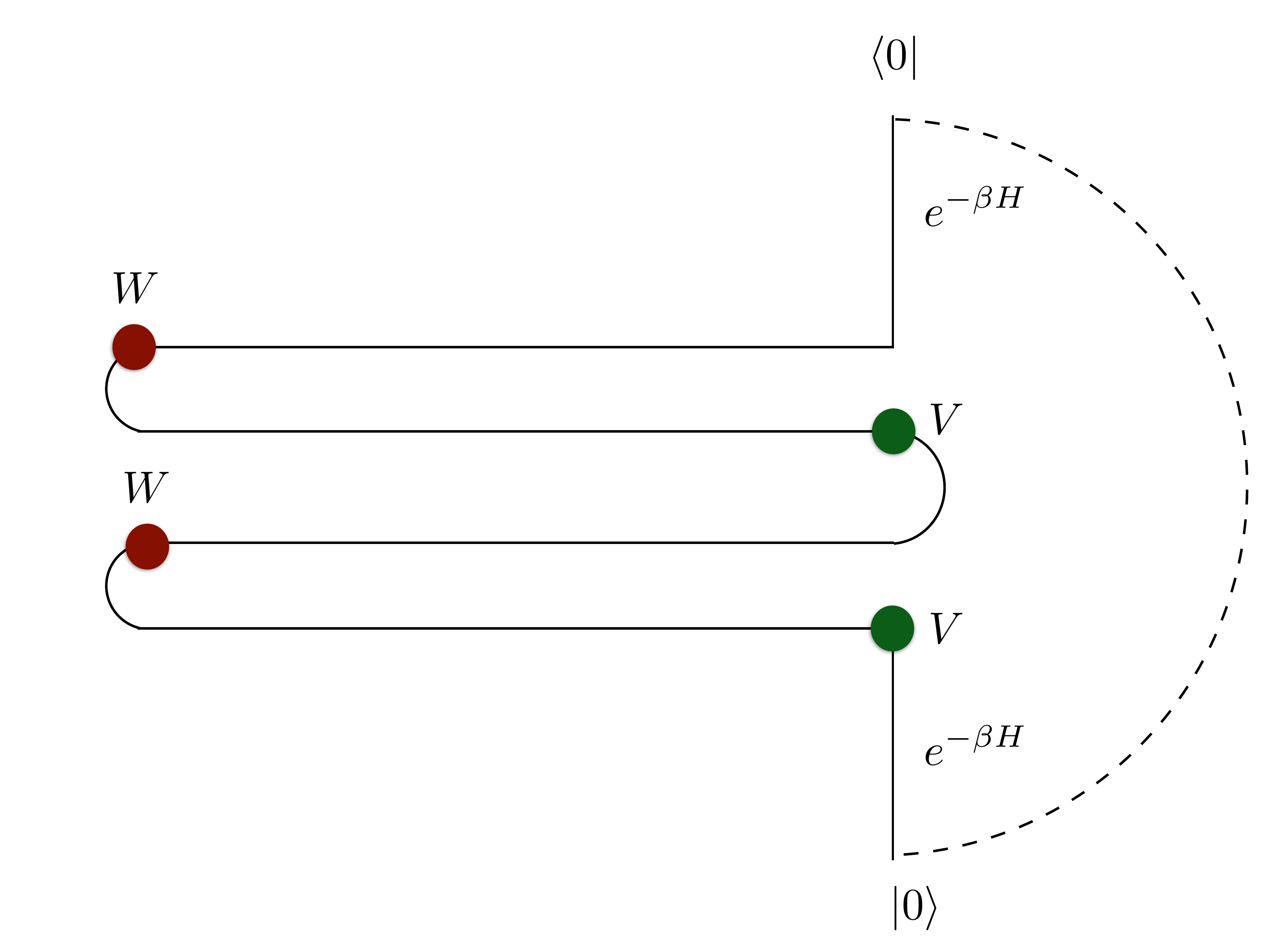}}
\caption{\small A demonstration of the Schwinger-Keldysh contour on which the four-point OTOC is calculated. The solid dots denote an insertion of the corresponding operator.} \label{otoc}
\end{center}
\end{figure}
In this article, we will specifically consider such $4$-point out-of-time-order correlator (OTOC). Such correlation functions are calculated on Schwinger-Keldysh contours shown in figure \ref{otoc}.\footnote{Strictly speaking, the bulk geometry for which one can construct the two-folded Schwinger-Keldysh contour is still not known explicitly. For our purpose, though, we will assume that such a description exists. We thank Gautam Mandal and Shiraz Minwalla for pointing this out and some discussion on related matters.} In this case, the correlator automatically becomes path ordered along the contour.

In our case, we will consider a specific class of probe degrees of freedom inside a bulk geometry. Thus, the corresponding Euclidean path integral takes the form:
\begin{eqnarray}
\Z_{\rm bulk} = \Z_{\rm gravity} \Z_{\rm probe} \equiv \Z_{\rm QFT} = \Z_{\rm bath} \Z_{\rm probe} \ .
\end{eqnarray}
Specifically, the QFT ``bath" degrees of freedom consists of adjoint matter and the ``probe" degrees of freedom consist of fundamental matter. In the bulk description, the $ \Z_{\rm probe}$ corresponds to a probe fundamental string or a D-brane and all fluctuations modes thereof. For a typical large $N_c$ gauge theory, we have 
\begin{eqnarray}
\Z_{\rm gravity} = \int {\cal D}[g] e^{-\frac{1}{G_N} S_{\rm sugra}} \equiv \Z_{\rm QFT} = \int {\cal D}[\phi] e^{- N_c^2 S_{\rm QFT}} \ , 
\end{eqnarray}
where, in the second line, we have converted the Newton's constant $G_N$ into the rank of the gauge group, $N_c$. One can certainly consider fluctuations of the ``gravity" modes and correspondingly an $n$-point correlation function, by expanding $S_{\rm sugra}$ around a suitable saddle point and expanding up to a desired order.

We will focus on a similar idea, on the probe sector, which is given by
\begin{eqnarray}
\Z_{\rm probe} = \int {\cal D}[\varphi] e^{- T_{\rm probe} S_{\rm probe}} \equiv \Z_{\rm QFT} = \int {\cal D} \left[\varphi^{\rm bdry} \right] e^{- N_c N_f S_{\rm QFT}} \ , 
\end{eqnarray}
where $T_{\rm probe}$ is the tension of the probe, $N_f$ is the corresponding number of fundamental matter field and $S_{\rm QFT}$ contains explicit factors of the 't Hooft coupling. We do not specify the integration measure in the functional integral. To compute a $4$-point correlation function, around a classical saddle, one simply expands the action upto quartic term and evaluates:
\begin{eqnarray}
\langle \cO_1 (x_1) \ldots \cO_4 (x_4) \rangle = - \left. \frac{\delta^4 S_{\rm QFT}^{\rm on-shell}} {\delta \varphi_1^{\rm bdry} (x_1) \ldots \delta \varphi_4^{\rm bdry} (x_4)} \right |_{\varphi_i^{\rm bdry} = 0 } \ , \label{probe4p}
\end{eqnarray}
where $S_{\rm QFT}^{\rm on-shell}$ is evaluated on the on-shell configuration of the quadratic fluctuations. Here the entire description is in the probe sector, while the background QFT degrees of freedom are merely spectators.

Given a classical profile for the probe sector, we can expand the action around this saddle, schematically:
\begin{eqnarray}
S_{\rm QFT} \left[ \{\varphi_i\} \right] =  S_{\rm QFT}^{(0)} \left[ 0 \right] + \frac{1}{2!} \varphi_i  \varphi_j \frac{\partial^2 S_{\rm QFT}^{(2)}}{\partial\varphi_i \partial \varphi_j }  + \frac{1}{4!} \varphi_i \varphi_j \varphi_k\varphi_l \frac{\partial^4 S_{\rm QFT}^{(4)}}{\partial\varphi_i \partial \varphi_j \partial \varphi_k \partial \varphi_l}  + \ldots  \ .
 \end{eqnarray}
Here $\{\varphi_i\}$ denotes general fields in the probe sector. To compute the four-point function in (\ref{probe4p}), we can simply find the saddle corresponding to $S_{\rm QFT}^{(2)}$ and evaluate $S_{\rm QFT}^{(4)}$, on-shell, with the data from the saddle point.

We will use the eikonal approximation to calculate the four point function, expressed as a $2-2$ scattering amplitude, pioneered in \cite{Shenker:2014cwa}. This approximation is valid for scattering processes with small angle and large incoming momenta. For our case, this holds true in the regime when $T_{\rm probe}$ is very large, with $T_{\rm probe} / s^{\alpha}$ held fixed.\footnote{In our case, $T_{\rm probe}$ plays the same role as the Newton's constant $G_N$ in \cite{Shenker:2014cwa}. } Here, $s$ is Mandelstam variable, corresponding to the scattering, and $\alpha$ is an appropriate number such that $T_{\rm probe} / s^{\alpha}$ becomes dimensionless.

This article is divided into the following sections: In section $2$, $3$ and $4$, we describe a classical string dynamics in an AdS$_3$-background and how one can already extract an effective Schwarzian action from the quadratic fluctuation analysis of the same. The next section is devoted to an explicit calculation of the Schwarzian sector with a generic fluctuation mode of the string, which is subsequently complemented by a path integral description of the effective description in the next section. Section $7$ discusses the sources of explicit conformal symmetry breaking. In the next section, an explicit example of semi-classical fluctuations is discussed, in which the large diffeomorphisms on the worldsheet has a particularly simple realization in terms of the fluctuations of the embedding coordinates. After concluding the discussion with fundamental strings, we move on to the dynamics of D-branes. This is done in section $9$, beginning with a discussion on the D$1$-brane. For our purposes, this is very similar to the discussion of fundamental strings, including the effective Schwarzian action. We subsequently discuss a probe sector of D$5$-branes, and obtain, in details, the Lyapunov exponent by calculating the two-two scattering amplitude of the D$5$-brane fluctuations. For analytical control, we discuss the gauge field fluctuations on this brane; in section $10$, the corresponding Lyapunov exponent is determined by the event horizon in the corresponding open string metric, and yields a non-vanishing value (with appropriately excited gauge field on the D-brane worldvolume) even when the background gravity degrees of freedom do not see any event horizon. In the next section, we present another analytically tractable example in which the $4$-point OTOC can be calculated: from the dynamics of a defect degree of freedom in a particular Lifshitz-type background. In section $12$, we demonstrate how a general functional of Schwarzian action can yield maximal chaos, and can violate the chaos bound otherwise. Finally, we conclude in section $12$, with some open problems and future directions. The main results of this article is summarized in \cite{Banerjee:2018twd}, here we discuss a more detailed account, including additional observations, comments and remarks.

\section{String in AdS$_3$-BH}

In this section, we will revisit the framework of \cite{deBoer:2017xdk},\footnote{See also \cite{Murata:2017rbp}.} with a different perspective.\footnote{Note that, in \cite{Cai:2017nwk} the presence of a Schwarzian effective action on a string worldsheet was also proposed. However, we have a direct and explicit description.} In particular, we will extract an effective Schwarzian action, starting from the Nambu-Goto action of the string, and studying the fluctuations thereafter. Towards that, we begin with the AdS$_3$-BTZ geometry, in the Fefferman-Graham patch:
\begin{eqnarray}
ds^2 = \ell^2 \left[ - \frac{1}{4} \left( r - \frac{r_{\rm H}^2 }{r}\right)^2  dt^2 + \frac{1}{4} \left( r + \frac{r_{\rm H}^2 }{r}\right)^2  dX^2 + \frac{dr^2}{r^2} \right]  \ . \label{btzfg}
\end{eqnarray}
In the above, $r_{\rm H}$ denotes the location of the event horizon, which sets the temperature: 
\begin{eqnarray}
T = \frac{r_{\rm H}}{2\pi} \ ,
\end{eqnarray}
and $\ell$ denotes the curvature scale: Ricci scalar, $\R = - 6/ \ell^2$.

It is straightforward to check that under the following map
\begin{eqnarray}
r = e^{1/\ell} \left( R + \sqrt{R^2 - R_{\rm H}^2 }\right) \ , 
\end{eqnarray}
the FG-metric in (\ref{btzfg}) maps to the standard Poincar\'{e} patch BTZ, given by
\begin{eqnarray}
ds^2 = - R^2 \left( 1 - \frac{R_{\rm H}^2}{R^2} \right) dt^2 + R^2 dX^2 + \left( 1 - \frac{R_{\rm H}^2}{R^2} \right)^{-1} \frac{dR^2}{R^2} \ . 
\end{eqnarray}
We will, however, work with the FG-patch description.

Similar to \cite{deBoer:2017xdk}, we consider a string, which is described by a Nambu-Goto dynamics, stretched in the AdS-radial direction. The end point of the string describes the trajectory of a fundamental degree of freedom ({\it e.g.}~a quark) in the thermal background of a large $N_c$ gauge theory. To describe this string, we need to solve an embedding problem. 

In general, we need a map, denoted by $\Phi$, describing a hypersurface $\Sigma_{1+1}$ into a three dimensional pseudo-Riemanian manifold $\cM_{2+1}$:
\begin{eqnarray}
\Phi : \quad \Sigma \hookrightarrow \cM \ .
\end{eqnarray}
Here $\cM_{2+1}$ is described by the FG-patch metric in (\ref{btzfg}), and we choose the embedding function $\Phi \equiv X\left( \tau, \sigma\right)$, where $\{\tau, \sigma\}$ define a patch for $\Sigma_{1+1}$, the string worldsheet. Furthermore, let us choose the static gauge for the hypersurface:
\begin{eqnarray}
\tau= t \ , \quad \sigma = r \ .
\end{eqnarray}

The dynamics of the string is described by the Nambu-Goto action:
\begin{eqnarray}
S_{\rm NG} = - \frac{1}{2\pi\alpha'} \int d\tau d\sigma \sqrt{- {\rm det} \gamma} \ , \quad \gamma_{ab} = G_{\mu\nu} \partial_a X^\mu \partial_b X^\nu \ ,
\end{eqnarray}
where $g_{\mu\nu}$ is the FG-patch metric given in (\ref{btzfg}) and $\alpha'$ is the string tension. The resulting equation of motion is given by
\begin{eqnarray}
\partial_a \left[\sqrt{- {\rm det} \gamma} \gamma^{ab} \left( \partial_b X^{\mu} \right) G_{\mu\nu}\left( X \right)  \right] - \frac{1}{2} \sqrt{- {\rm det} \gamma} \, \gamma^{ab} \left( \partial_\nu G_{\rho\alpha} \left( X \right) \right) \left( \partial_a X^\rho \right) \left( \partial_b X^\alpha \right) = 0 \ . \nonumber\\ \label{ngeom}
\end{eqnarray}
In deriving the above equation of motion, we view $G_{\mu\nu} \left( X \right)$ as a collection of the embedding functions $X(\sigma, \tau)$ and therefore explicitly included the variation of this collection of embedding functions, themselves. Alternatively, starting from the Polyakov action:
\begin{eqnarray}
S_{\rm Pol} = -  \frac{1}{4\pi\alpha'} \int d\tau d\sigma \sqrt{- {\rm det} h} \, h^{ab} G_{\mu\nu}\left( X \right) \left( \partial_a X^\mu \right) \left( \partial_b X^\nu \right)  \ ,
\end{eqnarray}
one arrives at the following equations:
\begin{eqnarray}
&& \partial_a \left[\sqrt{- {\rm det} h} \, h^{ab} \left( \partial_b X^{\mu} \right) G_{\mu\nu}  \right] - \frac{1}{2} \sqrt{- {\rm det} h} \, h^{ab} \left( \partial_\nu G_{\rho\alpha} \right) \left( \partial_a X^\rho \right) \left( \partial_b X^\alpha \right) = 0 \ , \\
&& T_{ab} = G_{\mu\nu} \left[ \left( \partial_a X^\mu \right) \left( \partial_b X^\nu \right) - \frac{1}{2} h_{ab} h^{cd} \left( \partial_c X^\mu \right) \left( \partial_d X^\nu \right) \right] = 0 \ . \label{tabpol}
\end{eqnarray}
Here $h_{ab}$ is the worldsheet metric, which, from (\ref{tabpol}), is obtained to be $h_{ab} = G_{\mu\nu} \left( \partial_a X^\mu \right) \left( \partial_b X^\nu \right) $. Therefore, as expected, $h_{ab} \equiv \gamma_{ab}$ in the NG-action. We will discuss the NG-action.

The simplest solution of the embedding function in (\ref{ngeom}) is given by $X\left( \tau, \sigma\right) \equiv X\left( t, r\right) = 0$. The corresponding worldsheet metric is obtained to be:
\begin{eqnarray}
ds_{\Sigma}^2 = \ell^2 \left[ - \frac{1}{4} \left( r - \frac{r_{\rm H}^2}{r}\right)^2  dt^2  + \frac{dr^2}{r^2} \right]  \ . \label{ads2fg}
\end{eqnarray}
It is straightforward to observe that (\ref{ads2fg}) is simply an AdS$_2$ in the corresponding FG-patch, whose curvature is given by $\R_{(2)} = - 2 \ell^2$.

Now, similar to \cite{deBoer:2017xdk}, we can consider fluctuations of the embedding function around the classical profile:\footnote{Note that, we will only consider transverse fluctuations here. Longitudinal fluctuations, {\it i.e.}~fluctuations in the $\{t, r\}$-directions in our static gauge, can not be undone by a worldsheet diffeomorphism, once we have chosen a gauge. However, we have not considered those.}
\begin{eqnarray}
X\left( t, r \right) = X_0 \left( t, r \right) + \delta X\left( t, r \right) \ ,
\end{eqnarray}
which yields:
\begin{eqnarray}
\gamma_{ab} = \gamma_{ab}^{(0)} \left[ X_{(0)}\right] + \delta \gamma_{ab} \left[ \delta X \right] \ . \label{metchange}
\end{eqnarray}
In the above, the superscript $(0)$ denotes the quantities corresponding to the classical profile of the string. Under an arbitrary fluctuation $\delta X\left( t, r \right) $, the change in the worldsheet metric $\delta \gamma_{ab}$ does not preserve the corresponding two-dimensional Ricci-scalar; hence, a general off-shell fluctuation can not be regarded as a diffeomorphism on the worldsheet. Here, we want to motivate that under certain conditions, the transverse fluctuations can indeed be considered as diffeomorphisms.

The dynamics of the fluctuation fields are governed by the Nambu-Goto action, expanded in fluctuation fields upto the quadratic order:
\begin{eqnarray}
S_{\rm NG} \left[ X \right] = S_{\rm NG}^{(0)} \left[ X_{(0)} \right] + S_{\rm NG}^{(2)} \left[ \delta X \right] \ .
\end{eqnarray}
The equation of motion for the fluctuation fields are obtained as:
\begin{eqnarray}
\delta S_{\rm NG}^{(2)} \left[ \delta X \right] = 0 \ . \label{eomfluc}
\end{eqnarray}
Now, we can attempt a general solution of (\ref{eomfluc}), in an FG-expansion. Towards that, we substitute an ansatz of the following form, in the equations of motion:
\begin{eqnarray}
\delta X \left(t, r \right) = r^\Delta \sum_{n=0}^{\infty} \frac{\delta X^{(n)} (t)}{r^{n}} \ . \label{anfg}
\end{eqnarray}
The indicial equation yields the following two roots for the exponent:
\begin{eqnarray}
\Delta = 0 \ , \, -3 \ . \label{delta}
\end{eqnarray}
Thus, we expect, in the series expansion, the field $\delta X$ can be solved up to two arbitrary functions of time: $\delta X^{(0)} (t)$ and $\delta X^{(3)} (t)$. This can be explicitly checked by substituting (\ref{anfg}) in (\ref{eomfluc}), which yields:
\begin{eqnarray}
\A \left[ \delta X^{(0)}, \delta X^{(3)}, \delta X^{(n)} \right ] = 0 \ ,  \quad n \not = 0, -3 \ , \label{algsol}
\end{eqnarray}
where $\A$, collectively, represents algebraic expressions involving the argument fields. Note that, viewed as a scalar field on the worldsheet, the UV behaviour of the fluctuation preserves the asymptotic data provided by the classical profile. Thus, from a UV CFT-perspective, the fluctuations $\delta X(t, r)$ consist of a marginal (corresponding to $\Delta =0$) and a relevant (corresponding to $\Delta = -3$) deformation.

One can explicitly check, for any value of $n\not = 0, -3$, equations (\ref{algsol}) can be solved explicitly:
\begin{eqnarray}
\delta X^{(n)} \left( t \right) = \F^{(n)} \left[ \delta X^{(0)}\left( t \right), \delta X^{(3)}\left( t \right)  \right] \ . \label{soleomfluc}
\end{eqnarray}
Here, the explicit form of $\F^{(n)}$ gets progressively involved, with increasing values of $n$. Thus, we refrain from explicitly presenting those.\footnote{Evidently, we cannot carry out this analysis for very large values of $n$, however, we have checked this claim explicitly to sufficiently high order of $n=30$. The pattern repeats itself for higher values, but explicitly, become cumbersome to keep track of. }

Now, we can evaluate the change in the two-dimensional Ricci-scalar $\delta\R_{(2)}$, as a function of $\delta \gamma_{ab}$. The generic structure of the Ricci-scalar, on-shell, takes the following form:
\begin{eqnarray}
\delta \R_{(2)} = r^\beta \sum_{n=0}^{\infty} \frac{\delta \R_{(2)}^{(n)} \left[ \delta X^{(0)}\left( t \right), \delta X^{(3)}\left( t \right) \right] }{r^{n}} \ , \label{riccifg}
\end{eqnarray}
where $\beta$ is some number which is unimportant for us. Now, it can be checked that, setting 
\begin{eqnarray}
\delta \R_{(2)} = 0 \quad \implies \quad \delta X^{(3)}\left( t \right) = - \frac{8}{3} \left[ r_{\rm H}^2 \left( \partial_t \delta X^{(0)}\left( t \right) \right) - \left( \partial_t^3 \delta X^{(0)}\left( t \right) \right) \right] \ . 
\end{eqnarray}
Therefore, Ricci-scalar preserving fluctuation is, ultimately, one function worth of freedom, denoted by $\delta X^{(0)}\left( t \right)$. If we hold the radial gauge fixed, {\it i.e.}~$\sigma = r$, then this one function worth of freedom is essentially the residual diffeomorphism, which also has a natural action at the conformal boundary. Thus, $\delta X^{(0)}\left( t \right)$ forms a representation of the Diff$(S^1)$ at the boundary. In this restricted sector, the physics of the fluctuations on the worldsheet can be viewed as a diffeomorphism invariant theory on the string worldsheet. In the next section we will discuss another such example.

\section{Strings in AdS$_3$}

Let us now consider the case where the string is embedded in a horizon-less background. Thus, we now consider an embedding problem in which $\cM_{2+1}$ is purely AdS$_3$, given by the following Poincar\'{e} patch
\begin{eqnarray}
ds^2 = \frac{R^2}{z^2} \left( - dt^2 + dX^2 + dz^2 \right) \ . \label{ads3Poincare}
\end{eqnarray}
We will discuss the case where the end point of the string undergoes an uniform acceleration. An explicit solution was obtained in \cite{Xiao:2008nr} and later analyzed in \cite{Jensen:2013ora}, specially in the context of the so-called ER=EPR conjecture. The embedding function is explicitly obtained as:
\begin{eqnarray}
X^{(0)} \left(t, z \right) = \pm \sqrt {a^{-2} + t^2 - z^2 } \ , \label{classacc}
\end{eqnarray}
where $a$ is the magnitude of the constant acceleration. The $\pm$ sign can be physically attributed to the identification of a quark or an anti-quark as the end point of the string.

The induced metric on the worldsheet is simply an AdS$_2$, which can be easily verified by calculating the two-dimensional Ricci-scalar: $\R_{(2)} = - 2/R^2$. The equations of motion for the fluctuations, $\delta X$, are given by
\begin{eqnarray}
\partial_t \left( \frac{\partial \sqrt{- {\rm det} \left( \gamma + \delta \gamma\right) }}{\partial (\partial_t \delta X)}\right) + \partial_z \left( \frac{\partial \sqrt{- {\rm det} \left( \gamma + \delta \gamma\right) }}{\partial (\partial_z \delta X)} \right) = 0 \ , 
\end{eqnarray}
The above equation may be solved for $\delta X\left( t, z\right)$, with an appropriate ansatz. However, for us, the algebraic expression of, {\it e.g.}~$\left( \partial_t^2 \delta X \right)$ in terms of $\delta X$ and its other derivatives is enough. Explicitly, this is given by
\begin{eqnarray}
&&\delta X^{(2,0)}(t,z) = \nonumber\\
&& -\frac{2 \left(a^{-2} + z^2\right) \delta X^{(0,1)}(t,z) + z \left(\left(z^2 - a^{-2} \right) \delta X^{(0,2)}(t,z) + 2  t \left(\delta X^{(1,0)}(t,z)+z \delta X^{(1,1)}(t,z)\right)\right)}{z \left(a^{-2} + t^2\right)} \ . \nonumber\\
\end{eqnarray}
Using the above equation of motion, it can now be explicitly checked that 
\begin{eqnarray}
\delta \R_{(2)} \left[ \delta X \right]  = 0 + \cO \left( \delta X\right)^3 \ .
\end{eqnarray}
Thus, the fluctuation modes, on-shell, around the classical solution in (\ref{classacc}) generate two-dimensional diffeomorphisms of the AdS$_2$ worldsheet.

\section{Schwarzian from Strings}

Given the seminal work by Polyakov in \cite{Polyakov:1987zb}, it is expected that an SL$(2,R)$ will play a pivotal role in describing the two-dimensional dynamics on the string worldsheet and therefore a Schwarzian effective action is also unsurprising. In this section, we will explicitly extract this Schwarzian sector. In the previous sections, we have explicitly demonstrated that, there exists a restricted set of worldsheet fluctuation modes on the worldsheet, which acts as a two-dimensional diffeomorphism on the worldsheet. Since the open string has a conformal boundary, it is natural to assume that there exists a natural sector of large diffeomorphisms as well. We will not belabour on the proof of an isomorphism between the diffeomorphism group and the set of worldsheet fluctuations. We will use this observation as a motivation to obtain the effective action of the large diffeomorphisms, based on the Nambu-Goto dynamics. In this respect, the presence of the conformal boundary breaks conformal invariance spontaneously, in which the large diffeomorphisms act as the Nambu-Goldstone bosons, similar to the ideas discussed in \cite{Maldacena:2016upp}.\footnote{See also \cite{Jensen:2016pah, Engelsoy:2016xyb}, in which the Schwarzian effective action is also obtained from slightly different perspectives.}

Towards that, we specifically seek an embedding of the string worldsheet in an asymptotically AdS$_3$ geometry. Generally, an asymptotically AdS$_3$ geometry is given by\cite{Roberts:2012aq}
\begin{eqnarray}
ds^2 = \ell^2 \left( L_+ dx_+^2 + L_- dx_-^2 - \left(\frac{2}{z^2} + \frac{z^2}{2} L_+ L_- \right) dx_+ dx_- + \frac{dz^2}{z^2} \right) \ , \label{ads3gen}
\end{eqnarray}
where $L_{\pm} \left( x_{\pm}\right) $ are two arbitrary smooth functions, which encodes the general diffeomorphism redundancy, after fixing the radial gauge. The patch in (\ref{ads3gen}) describes a locally AdS$_3$ geometry, with a non-trivial slicing. As is well-known, by selecting an appropriate slicing, one can reproduce the standard Poincar\'{e} AdS$_3$ patch in (\ref{ads3Poincare}) and the BTZ black hole in (\ref{btzfg}). Here, we will borrow the explicit results of \cite{Roberts:2012aq}, which maps Poincar\'{e} AdS$_3$ to the patch in (\ref{ads3gen}).

The explicit map is given by
\begin{eqnarray}
&& y_{\pm} = f_{\pm} \left( x_{\pm} \right) + \frac{2 z^2 f_{\pm}'^2 f_{\mp}''^2}{8 f_{\pm}' f_{\mp}' - z^2 f_{\pm}'' f_{\mp}''} \ , \label{Pointogen1} \\
&& u = z \frac{\left( 4 f_+' f_-' \right)^{3/2}}{ 8 f_{+}' f_{-}' - z^2 f_{+}'' f_{-}'' } \ , \label{Pointogen2}
\end{eqnarray}
which yields:
\begin{eqnarray}
ds^2 =\frac{ \ell^2}{u^2} \left( - 2 dy_+ dy_- + du^2 \right)  \ , \label{poinlc}
\end{eqnarray}
provided we identify:
\begin{eqnarray}
L_{\pm} = \frac{1}{2} \left\{ f_{\pm}, x_{\pm} \right \} = \frac{3 f_{\pm}''^2 - 2 f_{\pm}' f_{\pm}'''}{4 f_{\pm}'^2} \ . \label{schwarzian}
\end{eqnarray}
Thus, $L_{\pm}$ is precisely the Schwarzian derivative of the function $f_{\pm}$.  Also, the metric in (\ref{poinlc}) is the usual Poincar\'{e} slice of AdS$_3$, in the light-cone coordinates. Now, in the Poincar\'{e} patch of (\ref{poinlc}), we can begin by describing the simplest embedding of the string worldsheet.

Towards that, let us define:
\begin{eqnarray}
y_+ = \frac{1}{\sqrt{2}} \left( t + X\right) \ , \quad y_- = \frac{1}{\sqrt{2}} \left( t - X\right) \ .
\end{eqnarray}
This maps (\ref{poinlc}) to precisely (\ref{ads3Poincare}). The simplest embedding function is given by
\begin{eqnarray}
\Phi \equiv X \left( t, u \right) = 0 \ , \quad \implies \quad y_+ = y_- \ . 
\end{eqnarray}
This trivial embedding yields a worldsheet AdS$_2$. Thus, any worldsheet fluctuation, which is but a diffeomorphism, can simply be obtained by taking the $y_+ = y_-$ slice of the Poincar\'{e} section of the embedding AdS$_3$. We further assume that a natural action corresponding to the large diffeomorphism also exists. In the examples above, we have described a class of worldsheet embedding $\Sigma_{1+1}$, embedded in $\cM_{2+1}$. The explicit patches used to describe the manifold $\cM_{2+1}$ are simply diffeomorphic to each other.

Starting from (\ref{ads3gen}), we can easily take the $y_+ =  y_-$ slice. This slicing implies $x_+ = x_-$ and $f_{+} = f_{-}$, using (\ref{Pointogen1}) and therefore $L_+ = L_-$. Thus, on this slice, the metric on the string worldsheet is given by
\begin{eqnarray}
ds^2 = \ell^2 \left[ \frac{dz^2}{z^2} - \left( \frac{2}{z^2} + \frac{z^2}{2} L(x) ^2 - 2 L(x) \right) dx^2 \right] \ . 
\end{eqnarray}
The on-shell action coming from the Nambu-Goto action is therefore given by
\begin{eqnarray}
S_{\rm NG} = - \frac{1}{2\pi\alpha'} \int dz dx \frac{z^2 L(x)-2}{\sqrt{2} z^2} = - \frac{1}{2\pi\alpha'}  \int_{\partial {\rm AdS}_2} dx \frac{1}{2} \left( z L(x)+\frac{2}{z}  \right) \ . \label{ngos}
\end{eqnarray}
Recall that the conformal boundary is located at $z\to 0$, and thus, the on-shell quadratic action in (\ref{ngos}) contains a divergent piece and a vanishing contribution as $z\to 0$. We can introduce counter-terms to regulate the divergence at the UV. Note that, the $\partial {\rm AdS}_2$ contains two points: one at the UV and one at the IR. Let us denote these two points by $\epsilon_{\rm UV}$ and $\epsilon_{\rm IR}$.

Thus, the complete renormalized on-shell action is given by
\begin{eqnarray}
S_{\rm NG}^{({\rm ren})} = & - &  \frac{1}{2\pi\alpha'}   \int_{\partial {\rm AdS}_2} dx \frac{1}{2} \left( z  L(x) + \frac{2}{z}  \right) \nonumber\\
& + & \left. \frac{1}{2\pi\alpha'} \int dx \sqrt{\left( \frac{2}{z^2} + \frac{z^2}{2} L(x) ^2 - 2 L(x) \right)}  \right|_{z = \epsilon_{\rm UV} } \nonumber\\
= && \frac{1}{\sqrt{2} \pi \alpha'} \int dx \left( \frac{1}{\epsilon_{\rm IR}} + \frac{\epsilon_{\rm IR}}{2} L(x) \right)  + \cO \left( \epsilon_{\rm UV}^2 \right) \ .
\end{eqnarray}
Here, $\epsilon_{\rm IR}$ is a physical scale, such as the event horizon. The first term above is, therefore, simply a constant, which we can safely ignore for any dynamical aspect. The remaining piece in the effective action is therefore given by
\begin{eqnarray}
S_{\rm NG}^{({\rm ren})} = \frac{\epsilon_{\rm IR}}{2 \sqrt{2} \pi \alpha'} \int dx \left\{f(x) , x \right\} \ , \label{schnlin}
\end{eqnarray}
which is simply the Schwarzian action.

A few comments are in order: First, note that, keeping the $\epsilon_{\rm UV}$-term in the renormalized action, we will obtain a Schwarzian effective action which is supported by $\left( \epsilon_{\rm IR}- \epsilon_{\rm UV}\right) $, at the leading order in $\epsilon_{\rm UV}$. Here, evidently, $\epsilon_{\rm UV}$ is small, but $\epsilon_{\rm IR}$ is unconstrained. Setting $\epsilon_{\rm IR}=\infty$, we can still extract the effective Schwarzian action, provided we imagine a small but non-vanishing $\epsilon_{\rm UV}$. Similar to \cite{Maldacena:2016upp}, this $\epsilon_{\rm UV}$ can be thought of as the explicit cut-off which breaks conformal symmetry spontaneously. Therefore, depending on the situation, there may be physically meaningful appearance of the Schwarzian effective action at the UV or at the IR.

\section{Interacting System with the Schwarzian Sector}

\begin{figure}[ht!]
\begin{center}
{\includegraphics[width=0.8\textwidth]{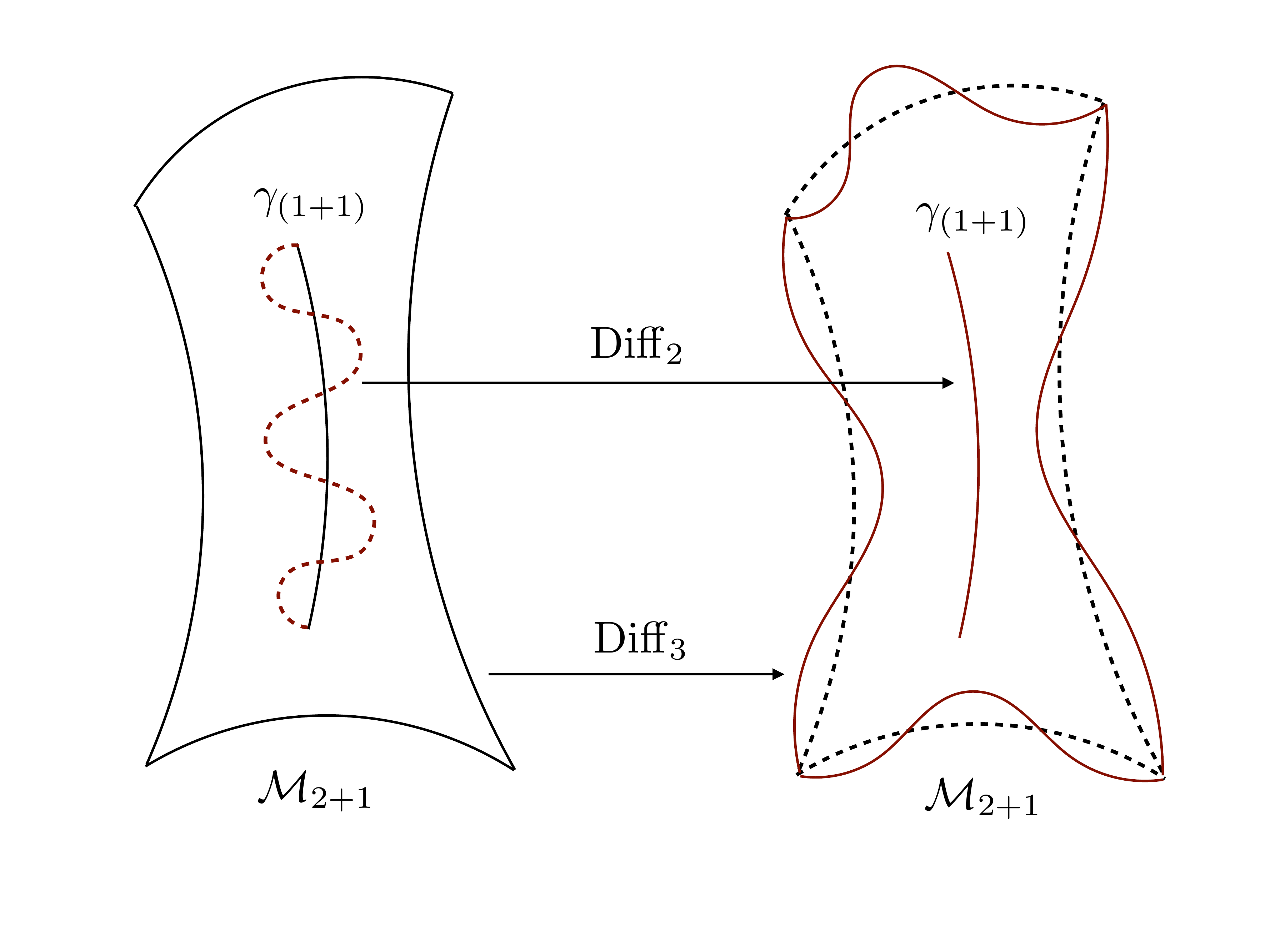}}
\caption{\small A cartoon demonstrating how Diff$_2$ can be obtained from a Diff$_3$. The left dashed curve represents a fluctuating worldsheet, embedded in an AdS$_3$. If the fluctuation belongs to the soft sector, then a Diff$_2$ can be used to map the fluctuating worldsheet back to the classical profile.} \label{sdiff}
\end{center}
\end{figure}

In this section, we will obtain the Schwarzian effective action, in its linearized form, and its' coupling to other open string modes, from a slightly different perspective. The basic idea is summarized in \cite{Banerjee:2018twd}, and here we elaborate on that. Here, we will consider transverse fluctuations only, however, one can carry out a similar analyses for longitudinal modes as well. In the previous section, we have demonstrated that imposing a rigid worldsheet condition ${\cal R}_{(2)} = -2$, {\it i.e.}~an exact AdS$_2$ worldsheet, does have non-trivial solutions for a fluctuating worldsheet as well.\footnote{In fact, we will see in a subsequent discussion that the rigid worldsheet condition is obeyed by either ingoing or outgoing fluctuation modes at the worldsheet event horizon, corresponding to the simple classical embedding $X=0$.} Therefore, physically, there exists an interesting and relevant class of transverse fluctuations which can be viewed as worldsheet diffeomorphisms, and therefore can be used to construct the large diffeomorphisms for the NG-theory. These diffeomorphisms can be easily constructed using the large diffeomorphisms of embedding AdS$_3$-background, as pictorially demonstrated in figure \ref{sdiff}. 

The NG-dynamics has the usual worldsheet diffeomorphism. In the context of holography, the worldsheet is a manifold with a (conformal) boundary, and therefore there is a natural notion of large diffeomorphism. Correspondingly, the large diffeos have a soft sector and we claim that this soft sector results in the maximal chaos, on the string worldsheet. Evidently, an arbitrary set of fluctuations around the $X=0$ classical saddle will not preserve the rigid worldsheet structure; moreover, an arbitrary classical profile itself may not yield an AdS$_2$ worldsheet. However, given any classical profile, the notion of a conformal boundary holds and therefore there is always a large diffeo on the worldsheet.

To make the most general claim about the nature of such a large diffeo, one needs to carry out a worldsheet analysis for an arbitrary classical embedding. This may be a difficult task and we will not pursue this here. Instead, we will take a simple case in which an explicit calculation can be performed. We will focus on the cases in which the classical embedding yields a worldsheet AdS$_2$, but we will not restrict the fluctuations.\footnote{Note that, this is a fairly generic class of string embeddings. The so-called Mikhailov embedding falls under this category.} Furthermore, we will work in the static gauge.\footnote{In principle, one should restrict a gauge condition for the complete analysis.}

Given this, the rational is simple: We use the standard Brown-Henneaux three dimensional large diffeos of the embedding AdS$_3$ background, and {\it project}\footnote{This projection is somewhat ambiguous, and we will elaborate more on this subsequently.} them on the AdS$_2$ worldsheet. Now, given any solution of the quadratic fluctuations around the AdS$_2$ string worldsheet,\footnote{Note that, for a large class of string embedding in AdS, the classical worldsheet is exactly AdS$_2$\cite{Mikhailov:2003er}. However, there are also physically very interesting configurations for which the classical worldsheet is no longer an AdS$_2$, {\it e.g.}~\cite{Gubser:2006bz}.} one can act by the large diffeos to create a new physics solution. This action of the large diffeos on the fluctuation solutions yield a coupling of the soft sector with an arbitrary fluctuation.

Towards that, in the FG-patch, the Brown-Henneaux diffeomorphisms are defined as:
\begin{eqnarray}
&& \xi^\mu \partial_\mu = r \xi^r \partial_r + \xi^t \partial_t + \xi^x \partial_x \ , \\
&& \xi^r = - \epsilon_+ '  - \epsilon_-'  \ ,  \\
&& \xi^t = \left( \epsilon_+ '  + \epsilon_-'  \right)  + \frac{2}{r^2}\left( \epsilon_+''   + \epsilon_-''  \right) + \frac{2 r_{\rm H}^2}{r^4}  \left( \epsilon_+''   + \epsilon_-''  \right) + \frac{2 r_{\rm H}^4}{r^6}  \left( \epsilon_+''   + \epsilon_-''  \right) \ldots \ , \\
&& \xi^x = \left( \epsilon_+ '  - \epsilon_-'  \right)  + \frac{2}{r^2}\left( \epsilon_+''   - \epsilon_-''  \right) + \frac{2 r_{\rm H}^2}{r^4}  \left( \epsilon_+''  - \epsilon_-''  \right) + \frac{2 r_{\rm H}^4}{r^6}  \left( \epsilon_+''  - \epsilon_-''  \right) \ldots \label{BrownHenneaux}
\end{eqnarray}
In the above, the $\pm$ sign corresponds to $(t \pm X)$-combinations. On the worldsheet, we can begin by projecting the Brown-Henneaux diffeomorphism on the worldsheet, using:
\begin{eqnarray}
\xi_a = \frac{\partial x^\mu}{\partial \sigma^a} \xi_\mu \ , \label{projection}
\end{eqnarray}
where $x^\mu$ are the embedding space coordinates and $\sigma^a$ are the worldsheet coordinates. If one takes a pull-back projection of $\xi^\mu$ on the worldsheet, we need to further consider $x^\mu$ with the embedding function data. Once we have constructed the two-dimensional analogue of Brown-Henneaux diffeomorphisms, we can calculate the change in the worldsheet metric with the action of a Lie derivative on the worldsheet metric itself:
\begin{eqnarray}
\gamma_{ab}^{\rm new} = \gamma_{ab}^{\rm old} + \delta \gamma_{ab} \ , \quad  \delta \gamma_{ab} = {\cal L}_{\xi} \gamma_{ab} \ .
\end{eqnarray}
To compute the interaction term, we can now evaluate the Nambu-Goto action, up to a certain order in the fluctuation.

Thus we evaluate:
\begin{eqnarray}
S_{\rm NG} \left[ \delta X^{(0)}(t), \epsilon(t) \right] = - \frac{1}{2\pi\alpha'} \left( \int_\Sigma dt dr  \sqrt{- {\rm det}\gamma_{ab}^{\rm new}} - \int_{\partial\Sigma} dt \sqrt{-{\rm det} h}\right) \ ,
\end{eqnarray}
where $\delta X^{(0)}$ is a non-normalizable (source) term of the fluctuation sector, see equation (\ref{anfg}), $\epsilon(t)$ denotes the soft sector coming from the large diffeos, $\Sigma$ represents the worldsheet manifold and $\partial\Sigma$ is the conformal boundary of the worldsheet; finally, $({\rm det} h)$ denotes the determinant of the boundary metric and this term is required for a UV-renormalization. On the worldsheet, $\epsilon_+$ and $\epsilon_-$ are locked together to produce a single field: $\epsilon(t) = \epsilon_+ = \epsilon_-$.

Let us evaluate the on-shell contribution in the Euclidean signature: $t \to - i \tau$. A direct calculation now yields:
\begin{eqnarray}
S_{\rm NG}^{(\rm E)} \left[ \delta X_{(0)}(\tau), \epsilon(\tau) \right] = S^{(\rm E)} \left[ \delta X_{(0)}(\tau) \right] + S^{(\rm E)} \left[ \epsilon(\tau) \right] + S_{\rm int}^{(\rm E)} \left[ \delta X_{(0)}(\tau), \epsilon(\tau) \right] \ ,
\end{eqnarray}
where
\begin{eqnarray}
S^{(\rm E)} \left[ \epsilon(\tau) \right] & = & c_1 \frac{\delta}{\pi\alpha'}\int d\tau \left[  \epsilon'(\tau) + \frac{\beta \epsilon'''(\tau)}{2 \pi } \right]  \ , \label{schinf} \\
 S^{(\rm E)} \left[ \delta X^{(0)}(\tau) \right] & = &  c_2 \frac{\delta^2}{\beta \alpha'} \int d\tau  \delta X_{(0)} '(\tau)^2   + \cO \left( \beta^0 \right) \ , \nonumber\\
S_{\rm int}^{(\rm E)} \left[ \delta X_{(0)}(\tau), \epsilon(\tau) \right] & = &  c_3 \frac{3\delta^3}{\pi \alpha'} \int d\tau  \left[ \delta X_{(0)}(\tau) \epsilon(\tau) \delta X_{(0)}'''(\tau )\right] \nonumber\\
& + & c_4  \frac{\delta^3}{\pi \alpha'} \int d\tau  \left[ \delta X_{(0)}'(\tau) \epsilon(\tau) \delta X_{(0)}''(\tau )\right] + \cO \left( \beta \right) \ .
\end{eqnarray}
Here $\{c_1, c_2, c_3, c_4\}$ are some numerical constants, which we do not keep track of. Furthermore, we have set the radius of AdS to unity. To obtain the above expressions, we have done the following: introduce a parameter $\delta$ which keeps track of the order of the fluctuation, which we will fix momentarily. Also, we have carried out a large temperature expansion, in which $\beta \ll 1$. In general, the above expressions receive corrections in powers of $\beta$.

Now, the Lie derivative action is a linear action and therefore we can only get a linear contribution in $\epsilon$. This is captured in (\ref{schinf}). Note that, even though this is a total derivative term and can be integrated exactly, the linear term consists of a third derivative in $\epsilon$, which is precisely what one expects from the linearization of a Schwarzian effective action, see {\it e.g.}~\cite{Maldacena:2016upp, Nayak:2018qej}. Thus, one expects that the non-linear completion of the $\epsilon'''(\tau)$ term yields the Schwarzian effective action. To make connection with (\ref{schnlin}), we choose:
\begin{eqnarray}
\epsilon_{\rm IR} = \frac{\sqrt{2}}{\pi} \frac{\delta} {\sqrt{\alpha'}} \ .
\end{eqnarray}

The second term above is purely a functional of the boundary field $\delta X_{(0)}(\tau)$. In the absence of $\epsilon$ modes, an $n$-point correlation function can be calculated by taking functional derivatives of this action, in the path integral description. To have a canonically normalized kinetic term, we can renormalize the fluctuation fields as in \cite{deBoer:2017xdk}, which amount to setting $\delta^2 \sim \alpha' \beta = \ell_s^2 \beta$, where $\ell_s$ is the string length. With this assignment, the Schwarzian action has a coupling constant $\sim \beta^{3/2}/\sqrt{\alpha'}$, the kinetic term has a coefficient $\sim \cO(1)$ and the interaction term has a coupling $\sim \beta^{3/2}\sqrt{\alpha'}$. Therefore, the scrambling time can be estimated by dividing the interaction coupling by the coefficient of the Schwarzian action. This yields: $\sim \alpha' $, which yields: $t_{\rm scr} \sim \beta \log \sqrt{\lambda}$, where $\lambda$ is the 't Hooft coupling. The assignment of $\delta^2 \sim \alpha'$ is therefore consistent with \cite{deBoer:2017xdk, Banerjee:2018twd}.

Finally, note that, as long as $\beta^{3/2} / \sqrt{\alpha'} \gg 1$, the Schwarzian action determines the saddle for $\epsilon(\tau)$, which is therefore the usual thermal saddle. The fluctuations $\delta X_{(0)}$ appear at $\cO(1)$ and are completely free at this order. Finally, the Schwarzian saddle configuration mixes with the fluctuation modes at the next order, $\cO(\sqrt{\alpha'})$, and in computing the four-point correlator of a generic fluctuation mode, the Schwarzian propagator for $\epsilon(\tau)$ can be used. This structure is very similar to what is discussed in \cite{Maldacena:2016upp}, for two-dimensional JT-gravity. 

There is another way in which one can deduce a 2D diffeomorphism on the worldsheet, given the Brown-Henneaux transformations for the target space AdS$_3$. In stead of (\ref{projection}) we simply find how the AdS$_3$ coordinates $\{r,t\}$ transform on the points on the worldsheet under the finite Brown-Henneaux transformations. This is done by substituting $x(r,t)$ corresponding to the static gauge worldsheet solution and not projecting the $\xi^x$ component in (\ref{BrownHenneaux}) onto the worldsheet. The finite version of such a 2D diffeomorphism has the following interpretation: we imagine that the fluctuating worldsheet has static coordinates which belong to a target space AdS$_3$. This is related to the target space of the non-fluctuating worldsheet by the finite version of (\ref{BrownHenneaux}). In such a case the actions are:
\begin{eqnarray}
S^{(\text{E})}[\epsilon(\tau)]&=&\frac{\delta}{\pi \alpha'}\int d\tau\,\left[-\epsilon'(\tau)+\frac{\beta}{2\pi}\epsilon'''(\tau)\right],\cr
S^{(\text{E})}_{\rm int}[\delta X_{(0)}(\tau),\epsilon(\tau)]& = &-\frac{\delta^3}{\beta \alpha'}\int d\tau\,\delta X'_{(0)}(\tau)\left[\delta X_{(0)}(\tau)\epsilon''(\tau)+\delta X'_{(0)}(\tau)\epsilon'(\tau)\right]\cr
&&\hspace{7.8cm}+\mathcal{O}(\beta)
\label{non_projection_action}
\end{eqnarray}    
where setting $\delta=l_s/r_H\sim\beta\sqrt{\alpha'}$ yields the scrambling time as observed in \cite{deBoer:2017xdk}.

\section{Path Integral Description}

Let us summarize the origin of the soft sector from an AdS/CFT perspective. Gauge-gravity duality enables us to write the the generating functional of correlators of the boundary CFT as a path integral in the bulk (super-)gravity action, in which the boundary values of the bulk fields, properly renormalized, act as sources in the boundary generating functional. In the large-$N$ limit$-$ $N$ being degrees of freedom  in the boundary theory; the leading order contribution comes from on-shell solutions in the bulk.

Let us consider the case where the bulk consists of a fundamental string embedded in an AdS-background (with or without a horizon). The dynamics is governed by the Nambu-Goto theory. This introduces a quark-like degree of freedom, at the conformal boundary of the embedding AdS-space. This fundamental matter is coupled to a large-$N$ Yang-Mills system. Now, one solves the equations of motion to obtain a classical profile of the string. This provides us with the classical saddle. To construct a semi-classical description, one now considers linearized fluctuations around the classical saddle.

In general, the fluctuations will have two distinct modes near the AdS boundary: non-normalizable and normalizable. The bulk solution is then taken as a linear combination of these two modes, while boundary value of the non-normalizable mode acts a source. The normalizable mode is fixed by demanding regularity of the fluctuation in the deep interior. Usually the specific condition one imposes for regularity depends on the the physics in question, but this, nonetheless, determines the normalizable mode in terms of the source.

The bulk Nambu-Goto action has a gauge symmetry associated with world-sheet diffeos. Since this is a (2d) gauge theory on a manifold with boundary, there can be relevant large gauge transformations to which one can associate a finite charge. We claim that this is the soft sector responsible for chaos, as seen in \cite{deBoer:2017xdk}. Further, the effective action associated with these  modes is that of a Schwarzian action (if euclideanized) defined on a thermal time circle of periodicity $\beta$ associated to the event horizon temperature of the worldsheet geometry. The method used in \cite{deBoer:2017xdk} employs eikonal approximation, using shock-waves to calculate the 4pt. function beyond the probe approximation. This must therefore be similar to computing the bulk on-shell action perturbatively in string tension beyond the linearised fluctuations \cite{Poojary:2018ronp}.

For a better analogy, we can contrast this case with that of a probe scalar field in AdS, as in \cite{Shenker:2014cwa}. There the scalar is coupled to gravity minimally. The first order in back-reaction of the scalar onto the background metric allowed for the scalar modes to couple to the the soft modes which existed in the gravitational perturbation around a specific black hole solution.

Here, however the background gravity is just a spectator and the entire dynamics is in the Nambu-Goto action\footnote{One can consider the Nambu-Goto string probing the background in a manner similar to \cite{Shenker:2014cwa}.}. Around a specific solution of the Nambu-Action, there would exist on-shell world-sheet fluctuations which solves the linearized NG $eom$. For every such fluctuation one can write another set of fluctuations which are generated by the large 2d diffeomorphisms acting on them. In other words, since diffeomorphisms generate new solutions of fluctuations, action of diffeomorphism on the fluctuations provides a coupling with the soft modes. Here, both the hard and soft sectors are therefore viewed as fluctuations on the world-sheet.

Since we discuss the derivation of the effective action associated with the soft modes, it would be useful to know how the eikonal approximation of \cite{deBoer:2017xdk} and \citep{Shenker:2014cwa} gets formally related to the path integral in the bulk. To explain this we take the case of the scalar field in the bulk \citep{Shenker:2014cwa,Poojary:2018ronp}. Let us consider a 4 pt. function of scalars in the bulk which are minimally coupled to gravity, following \cite{Kabat:1992tb}:
\begin{eqnarray}
&&G(x_1,x'_1 ; x_2,x'_2)=\cr
&&\,\,\int{\mathcal{D}[h]\mathcal{D}[\phi_1]\mathcal{D}[\phi_2]\,\,\phi_1(x_1)\phi_1(x'_1)\phi_2(x_2)\phi_2(x'_2)}\,\,{\rm exp}\left\lbrace i\int d^d x \sqrt{-{\rm det}g(h)}\right.\cr
&&\,\,\,\,\left.\left[-\frac{l}{16\pi G}(R(h)-\Lambda)-\frac{1}{2}\left(\nabla_\mu \phi_i\nabla^\mu \phi_i+m^2_i\phi_i\phi_i\right) \right]\right\rbrace \ .
\end{eqnarray}
We can latter take the the bulk points to the boundary, taking an appropriate limit should yield the corresponding boundary 4pt. function sourced by the boundary values of the bulk scalar. Since putting the scalars to zero is a consistent truncation of the this theory we will work around a solution with no scalars, and consider linearised fluctuations in the metric denoted by $h$. If we are interested in the 4pt. functions computed from diagrams which receive contribution only from graviton lines connecting the different scalar lines and not the same scalar, then we can write the above path integral as
\begin{equation}
\int{\mathcal{D}}[h]G_1^c(x_1,x'_1\vert h_{\mu\nu})G_2^c(x_2,x'_2\vert h_{\mu\nu})\,\,{\rm exp}\left\lbrace i\int d^dx \tfrac{1}{2}hD^2h\right\rbrace,
\label{eikonal2}
\end{equation}
where $G^c(x,x'\vert h_{\mu,\nu})$ is the connected part of the scalar propagator evaluated upto first order in metric perturbation $h_{\mu\nu}$. As demonstrated in Kabat and Ortiz\cite{Kabat:1992tb}, in the eikonal limit for massless particles the above corelation fuction reduces to:
\begin{equation}
\int {\mathcal{D}}[h]{\rm exp}\left\lbrace i\int d^dx\sqrt{-g}\left( \tfrac{1}{2}hD^2h + \tfrac{1}{2}h_{\mu\nu}T_1^{\mu\nu}+ \tfrac{1}{2}h_{\mu\nu}T_2^{\mu\nu}\right)\right\rbrace \ , 
\label{eikonal_4ptfunction}
\end{equation}
where $T_1^{\mu\nu}\&T_2^{\mu\nu}$ are stress-tensors for massless particles along their null trajectories. Evaluating the above path-integral on saddle point implies that $h_{\mu\nu}$ be the back-reaction due to the stress tensor. This is precisely the origin of the phase factor $e^{i\delta(s)}$ in \citep{Shenker:2014cwa,deBoer:2017xdk}.

Let us take step back to (\ref{eikonal2}), here itself we could have taken the limit where the bulk points $x_1,x_1',x_2,x_2'$ are taken to the boundary, then (\ref{eikonal2}) would look like:
\begin{equation}
\int{\mathcal{D}}[h]\mathcal{O}_1(x_1,x'_1\vert h_{\mu\nu})\mathcal{O}_2(x_2,x'_2\vert h_{\mu\nu})\,\,{\rm exp}\left\lbrace i\int d^dx \tfrac{1}{2}hD^2h\right\rbrace \ ,
\label{eikonal4}
\end{equation}
which is basically the next order in $G_N$ of the expression:
\begin{eqnarray}
&&\langle\mathcal{O}_1(x_1)\mathcal{O}_1(x_1')\mathcal{O}_2(x_2)\mathcal{O}_2(x_2')\rangle =\frac{\partial^2}{\partial \phi^{(0)2}_1}\frac{\partial^2}{\partial \phi^{(0)2}_2}Z(\phi^{(0)}_1,\phi^{(0)}_2)\vert_{\phi^{(0)}_1=\phi^{(0)}_2=0}\cr
&&Z(\phi^{(0)}_1,\phi^{(0)}_2)=\int \mathcal{D}[g]\mathcal{D}[\phi_i]\,\,{\rm exp}\left\lbrace\int \sqrt{-g}\left[\tfrac{l}{16\pi G_N}(R-2\Lambda)+(\partial \phi_i)^2-m_i^2\phi_i^2\right]\right\rbrace\cr&&
\end{eqnarray}
The above is evaluated in the probe approximation. In other words (\ref{eikonal4}) can be written as:
\begin{equation}
\int{\mathcal{D}}[g]\mathcal{O}_1(x_1,x'_1\vert g_{\mu\nu})\mathcal{O}_2(x_2,x'_2\vert g_{\mu\nu})\,\,{\rm exp}\left\lbrace \frac{il}{16\pi G_N}\int d^dx \sqrt{-g}(R-2\Lambda)\right\rbrace \ ,
\label{eikonal5}
\end{equation}
upto some overall constant.

So, to sum up, the soft sector can be recovered from the bulk on-shell action evaluated about a saddle point and constrained by the boundary conditions imposed on the metric path integral. In the case of AdS$_3$, absence of bulk propagating degrees of freedom implies that all on-shell metrics continuously connected to a saddle point geometry (ex. a black hole) are parametrized by the functions on the boundary. In this case, it would be the functions parametrizing the CFT$_2$ conformal transformations. Furthermore, $\mathcal{O}(x,x'\vert g_{\mu\nu})$ would correspond to knowing how the boundary CFT 2pt. function changes under conformal transformation, thus coupling to the action of the soft modes in the exponent. This would be similar to evaluating $Z(\phi^{(0)}_1,\phi^{(0)}_2)$ for a family of AdS$_3$ geometries about a given metric and reading off the exponential series in terms of the sources.

The procedure is the same when dealing with the string governed by the Nambu-Goto action in AdS$_3$\cite{Poojary:2018ronp}. The Brown-Henneaux large gauge diffeos in AdS$_3$ are the relevant large gauge transformations which are responsible for the soft sector. In the case of the NG string, it would be the large world-sheet diffeos. One therefore needs to evaluate the NG action for an arbitrary linearised fluctuation and for some relevant large gauge diffeos. These diffeos can also be thought of as further fluctuations on the world-sheet metric associated with a particular linearised fluctuations, but explicitly showing this might be cumbersome and not required.

The path integral perspective for the NG string is as follows: For small fluctuations of the string, the boundary $n$-pt functions are generated by the bulk on-shell path integral:
\begin{equation}
e^{S^{(0)}_{\rm NG}+S^{(2)}_{\rm NG}[x^{(0)}]} = Z_{\rm CFT}[J] \vert_{J=x^{(0)}} \ ,
\end{equation}    
where $S^{(0)}_{\rm NG}$ is the renormalized on-shell action for the solution to NG $eom$ about which linearizsed fluctuations exist. The later are captured in $S^{(2)}_{\rm NG}$, which is the on-shell quadratic action with boundary value $x^{(0)}(t)$. The effect of the soft modes is captured by doing large 2d diffeos on this configuration and re-evaluating the on-shell action, the path integral then becomes:
\begin{equation}
\int {\mathcal{D}}[f]\,\,e^{S^{(0)}_{\rm NG}+S^{(2)}_{\rm NG}(x^{(0)},f)+S_{f}} = Z_{\rm CFT}[J] \vert_{J=x^{(0)}} \ ,
\end{equation}    
where $f(t)$ parametrizes the large diffeomorphisms, $S^{(2)}_{\rm NG}(x^{(0)},f)$ captures the change due to this and $S_f$ is the action associated with the soft modes. Notice that the above path integral in $f$ translates to computing $n$-pt functions by differentiating $w.r.t.$ $x^{(0)}$ and contracting bilinears in $f$, using the propagator from $S_f$ as in the section-4.2 of \cite{Maldacena:2016upp}. Once one finds the effective action for the soft modes and how the 2pt. functions couple to them, once can work in an Euclidean setting and use the $i\epsilon$  prescription to get the required OTOC \citep{Maldacena:2015waa}. We will also explicitly carry out a similar computation in a later section.

\section{Explicit Breaking of Conformal Symmetry}

Let us begin commenting on the backreaction of the fluctuations on the classical string worldsheet. As we have described in (\ref{anfg})-(\ref{delta}), the fluctuations source a marginal and a relevant deformation at the UV boundary. Therefore, it is reasonable to assume that the classical embedding receives vanishing contribution from the fluctuation backreaction. Now, at the IR, {\it e.g.}~near $r = r_{\rm H}$, the fluctuation equation for $\delta X(t,r)$ can be similarly solved in a Frobenius series:
\begin{eqnarray}
\delta X (t, r) = \left( r - r_{\rm H}\right)^{\Delta_{\rm IR}} \sum_{n=0}^{\infty} \delta X_{\rm IR}^{(n)} (t) \left( r - r_{\rm H } \right) ^{n} \ , \label{anfgIR}
\end{eqnarray}
where $\delta X_{\rm IR}^{(n)}$ can be determined order by order. The indicial equation involves $\delta X_{\rm IR}^{(0)}$ directly, which can be solved with an oscillatory ansatz: $\delta X_{\rm IR}^{(n)} \sim e^{i \omega t}$. The indicial equation now yields:
\begin{eqnarray}
\Delta_{\rm IR}^2 + \frac{\omega^2}{4 r_{\rm H}^2} = 0 \ , \, 1 \ . \label{deltair}
\end{eqnarray}
Thus, there are two modes in the IR, with opposite signs. Depending on the spectrum of the theory, one obtains $\Delta_{\rm IR}$, which can have a growing mode, a decaying mode or a marginal mode towards the UV.

On the other hand, the fluctuation modes grow towards the IR, as seen from the UV-perspective. In principle, this entire system is well-defined and closed, without invoking the need of any new physics either at the UV or at the IR. Na\'{i}vely, the fluctuation backreaction can also be consistently ignored, both at the UV and at the IR. However, this is not precisely correct and we will make this statement more quantitative momentarily. Therefore, it remains unclear how conformal symmetry is broken in the framework.

Before moving further, let us discuss the behaviour of $\delta X$, in terms of its near boundary and near horizon behaviour, when the embedding space is an AdS$_{d+1}$. Carrying out an expansion similar to (\ref{anfg}), we recover $\Delta = 0 , -3$ for the conformal boundary. Similarly, the IR expansion of (\ref{anfgIR}) yields 
\begin{eqnarray}
\Delta_{\rm IR} = -  \frac{\omega^2}{d^2 r_{\rm H}^2} \ .
\end{eqnarray}
These data are independent of the embedding AdS$_{d+1}$-background, except for the overall constant. As before, knowing the complete theory we can identify the nature of this mode at the IR.

We will offer a simple picture of the breaking of conformal invariance. We have considered the open string degrees of freedom as probes in a given supergravity geometry. Let us assume that we do not introduce any more stringy degrees of freedom (such as a D-brane, located at the UV, on which the string ends). Even in this case, the open strings will backreact on the classical supergravity geometry. In fact, it is known that backreaction of such open strings are not negligible in the deep IR, and they change the IR physics qualitatively\cite{Kumar:2012ui, Faedo:2014ana}. Let us imagine the AdS$_3$-background as the near-horizon geometry of a D$1$-D$5$ brane bound state. The corresponding supergravity solution is characterized by a metric, a constant dilaton and a three form flux. Explicitly, these data can be given by
\begin{eqnarray}
&& ds^2 = \frac{r^2}{L^2} \left( - dt^2 + dx^2 \right)  + \frac{L^2 dr^2}{r^2} + L^2 d\Omega_3^2 + \left( \frac{Q_1}{Q_5} \right)^{1/2} ds_{{\cal M}_4}^2 \ , \label{ads3sugra} \\
&& G_3 = 2 Q_5 \left( \epsilon_3 + i *_6 \epsilon_3 \right)  \ , \\
&& e^{-2\phi} = \frac{Q_5}{Q_1} \ , \quad L^4 = Q_1 Q_5 \ .
\end{eqnarray}
Here $d\Omega_3^2$ is the line element along an $S^3$, $ds_{{\cal M}_4}^2$ is the line element along a four-manifold which is, typically, a $T^4$ or the $K3$-manifold. The volume form of the $S^3$ is represented by $\epsilon_3$ and $*_6$ is the Hodge dual operation defined on the six-dimensional manifold: AdS$_3 \times S^3$.

Given the background in (\ref{ads3sugra}), we can compare the Einstein tensor of the geometry with the stress tensor of an open string with a worldsheet along the $\{t, r\}$-plane. For this, it is enough to consider the Nambu-Goto action, and vary the action with respect to the background metric. This yields:
\begin{eqnarray}
T_{\rm strings}^{\mu\nu} = \frac{2\kappa_{10}^2}{2\pi\alpha' \sqrt{- G}}  \frac{\delta S_{\rm NG}}{\delta G_{\mu\nu}} = \frac{ \kappa_{10}^2}{ \pi\alpha' \sqrt{- G}} \sqrt{- \gamma} \gamma^{ab} \delta_a^\mu \delta_b^\nu \ , \label{tmnstring}
\end{eqnarray}
where $\kappa_{10}^2$ is the ten-dimensional Newton's constant, $\gamma$ is the induced worldsheet and $G$ is the background metric. The background Einsten tensor and the string stress tensor are given by
\begin{eqnarray}
E_{tt}^{\rm background} \sim \frac{r^2}{L^4} \ , \quad T_{tt}^{\rm string} \sim \frac{r \kappa_{10}^2 }{4 L \pi \alpha'} \ . \label{ETcom}
\end{eqnarray}
Clearly, in the $r \to 0$ limit, $T_{tt}^{\rm string}$ begins to dominate over $E_{tt}^{\rm background}$. In fact equating $E_{tt}^{\rm background}$ with $T_{tt}^{\rm string}$ we can derive a radial scale, which is the lower bound for the probe approximation to remain valid:
\begin{eqnarray}
r_{\rm IR} = \frac{L^3 \kappa_{10}^2}{\pi \alpha'} \ . \label{rIR}
\end{eqnarray}
Beyond $r_{\rm IR}$, the string backreaction modifies the IR into a Lifshitz-symmetric background\cite{Faedo:2014ana}, in which the dilaton field acquires a non-trivial profile and breaks the conformal symmetry explicitly. Therefore, a natural radial IR cut-off is an event horizon $r_{\rm H} \ge r_{\rm IR}$, which, qualitatively, signifies the presence of the explicit symmetry breaking.

In fact, embedding an open string in a general AdS$_{d+1}$-background yields identical results as compared to (\ref{ETcom}) and (\ref{rIR}), except of overall numerical constants. Taking the AdS$_5$-background, in which we further identify:
\begin{eqnarray}
&& L^4 = 4 \pi g_s N_c \alpha'^2 \ , \quad \lambda = g_{\rm YM}^2 N_c = 4 \pi g_s N_c \ , \\
&& \kappa_{10}^2 \sim  g_s^2 \alpha'^4 \ . 
\end{eqnarray}
In this case, we get:
\begin{eqnarray}
r_{\rm IR} \propto \frac{\sqrt{\lambda}}{16 \pi^2 N_c^2 } \ , \quad {\rm with} \quad L^4 = 1 \ . \label{rIRgauge}
\end{eqnarray}
Finally, using (\ref{tmnstring}), we can evaluate the stress-tensor of the fluctuation sector, as compared to the classical saddle configuration. In the BTZ background, around the simple $X=0$ classical saddle, the stress-tensor behaves as:
\begin{eqnarray}
T_{tt}^{\rm string} \sim \frac{r^4}{4 \left( r^2 + r_{\rm H}^2 \right)^2 } + \frac{r^4}{4 \left( r^2 - r_{\rm H}^2 \right)^2 }\left( \partial_t \delta X\right)^2 \ .
\end{eqnarray}
Clearly, the fluctuation stress tensor grows towards the IR and as $r \to r_{\rm H}$, the second term above starts dominating. Equivalently, the classical profile start receiving backreaction from the fluctuation sector itself. Note that, there is still a limit of validity for a perturbative fluctuation analysis, as long as the energy associated with the fluctuation sector is small enough. For example, if we assume a simple oscillatory behaviour for $\delta X$ in real time, {\it i.e.}~$\delta X \sim e^{i \omega t}$, the perturbative analysis remains valid provided:
\begin{eqnarray}
r^2 - r_{\rm H}^2 \ll 1 \ , \quad \omega \ll 1 \ , \quad {\rm with} \quad  \frac{\omega^2}{\left( r^2 - r_{\rm H}^2 \right)^2} \ll 1 \ . 
\end{eqnarray}
The frequency $\omega$ can be tuned accordingly, provided there is no gap in the spectrum.

Let us now consider the same question in an AdS$_{d+1}$ background:
\begin{eqnarray}
ds^2 = - r^2 f(r)  dt^2 + r^2 d\vec{x}_d^2 + \frac{dr^2}{r^2 f(r) } \ , \quad f(r) = 1- \frac{r_{\rm H}^{d}}{r^{d}}  \ .
\end{eqnarray}
A similar calculation yields the following stress-tensor component:
\begin{eqnarray}
T_{tt}^{\rm string} \sim \frac{r^{7-3d}}{r_{\rm H}^2 \left( r^4 - r_{\rm H}^4 \right) } \left( r^d r_{\rm H} - r_{\rm H}^d r \right)^2 \left[ 1 + \frac{r^4}{ \left( r^4 - r_{\rm H}^4 \right)} \left( \partial_t \delta X\right)^2 \right] \ .
\end{eqnarray}
Here $\delta X$ is the corresponding fluctuation. Clearly, near the horizon $r \to r_{\rm H}$, the fluctuation backreaction increases. As before, on the class of oscillatory (in time) solutions for $\delta X$ can be analyzed purely within the perturbative regime, with an appropriate tuning of the frequency. Note that, in the above expressions, $r$, $r_{\rm H}$ etc are dimensionless\footnote{We have further set the AdS curvature scale to unity.} and $\omega$ is measured in units of the temperature. The denominator introduces a natural IR cut-off, denoted by $\delta = r - r_{\rm H}$. For the perturbative treatment to hold, we must impose: $\omega^2 r_{\rm H}^4 \ll \delta $. Here, $\delta$ is necessarily non-vanishing, and therefore it introduces a stretched horizon above the event horizon. This observation was already made in \cite{deBoer:2008gu, Son:2009vu} and a {\it membrane-paradigm} like interpretation was advocated.

Before leaving this section, we will offer a few comments on the UV nature of this system. The open string has an UV end point and presumably ends on a D-brane. The complete description of this system is based on the work of \cite{Karch:2002sh}. A typical such D-brane construction is made of $N_c$ number of D$3$-branes and an $N_f$ number of D$7$-branes. In this case, the string end point at the UV ends on the D$7$ brane and, in turn, the proper action of the string represents the mass of the matter field that is introduced in the dual gauge theory. The UV completion of such models can be rather rich and involved, including the generic possibility of the breaking of conformal symmetry. This is similar to conformal symmetry breaking in the UV of a two-dimensional dilaton gravity theory. Because of the possible backreaction of various fields, a similar description to the dilaton gravity model may exist, which we will not further explore in this article. Instead, as in \cite{deBoer:2008gu}, one can simply impose a Neumann boundary condition at the string end point at UV.

\section{Fluctuations on the Worldsheet: Explicit Example}

Generically, the embedded string in an AdS$_{d+1}$ background has a simple classical saddle denoted by $X = 0$. Around this saddle, the semi-classical fluctuation analyses yields the following action:
\begin{eqnarray}
S_{(2)} &  = & \frac{1}{4\pi\alpha'} \int d^2 \sigma G_{IJ} \sqrt{- {\rm det}\gamma} \gamma^{ab} \left( \partial_a \delta X^I \right) \left( \partial_b \delta X^J \right) \ , \label{NGtwo} \\
& = & - \frac{1}{4\pi\alpha'} \int d^2\sigma \left( {\rm EOM} \right)  + \frac{1}{4\pi\alpha'} \int_{\partial\Sigma} d\sigma G_{IJ} \sqrt{- {\rm det}\gamma} \gamma^{ar} \left( \partial_a \delta X^I \right) \delta X^J \label{NGtwo1}
\end{eqnarray}
where $I, J$ denote the transverse directions. For AdS$_3$ background, $G_{IJ}$ contains only one component. The variation of (\ref{NGtwo}) yields the following equation of motion:
\begin{eqnarray}
{\rm EOM} = \frac{1}{\sqrt{- {\rm det}\gamma}} \partial_a \left[ \sqrt{- {\rm det}\gamma} \gamma^{ab} G_{IJ} \partial_b \delta X^J \right] = 0 \ . \label{eomfluc}
\end{eqnarray}
The fluctuations can be analytically solved in AdS$_{3}$-background\cite{deBoer:2008gu}. In this section we will make direct use of the analytical solution to discuss the effective Schwarzian action and its coupling to other modes in the theory.

It's particularly simple to work in the Poincar\'{e} patch, in which the general solution of (\ref{eomfluc}) is given by\cite{deBoer:2008gu}
\begin{eqnarray}
\delta X (t, r) = \frac{c_1 (r-i \omega )}{r} \left(\frac{r - r_{\rm H}}{r + r_{\rm H}} \right)^{ - \frac{i \omega }{2 r_{\rm H}}} + \frac{c_2 ( r + i \omega )}{r} \left(\frac{r- r_{\rm H}}{r + r_{\rm H}}\right)^{\frac{i \omega }{2 r_{\rm H} }} \ . \label{delX3}
\end{eqnarray}
In the above solution, both incoming and outgoing modes near the event horizon has been kept, and, so far, no boundary condition has been imposed. In keeping with the asymptotic expansion near conformal boundary, it is easy to check that the asymptotic expansion of (\ref{delX3}) takes the form:
\begin{eqnarray}
\delta X(t, r) = \left( c_1 + c_2 \right) e^{i \omega t } + \frac{c_1 + c_2 }{2} \frac{\omega^2}{r^2} e^{i \omega t} + i \frac{c_1 - c_2}{3} \frac{\omega\left( \omega^2 + r_{\rm H}^2\right) }{r^3} e^{i \omega t } + \ldots 
\end{eqnarray}
Thus, the two independent modes correspond to $\Delta = 0 , -3$. Evaluating the boundary term in (\ref{NGtwo1}) yields:
\begin{eqnarray}
S_{(2)} = \frac{1}{4\pi\alpha'} \int dt e^{2i\omega t} \left( c_1 + c_2 \right) \left( c_1 - c_2 \right) \ , 
\end{eqnarray}
which can be naturally interpreted as the deformation in the boundary theory. Defined in terms of (\ref{NGtwo1}), both $\Delta = 0, -3$ modes are normalizable since they yield a finite action.

The string has two end points: one at the IR event horizon and one at the UV conformal boundary. To account for a finite mass of the open string degree of freedom, one can impose Neumann boundary condition at the UV, as in \cite{deBoer:2008gu}: $\partial_r \delta X = 0 $ at $r=r_{\rm UV}$. The cut-off $r=r_{\rm UV}$ can be identified with the bare mass of the matter sector, by simply evaluating the Nambu-Goto action of a rigid string extended between $r_{\rm H}$ and $r_{\rm UV}$. This yields\cite{deBoer:2008gu}: $m \sim r_{\rm UV} / (\alpha' \beta)$, where $\beta$ is the period of the Euclidean time.

The Neumann boundary condition leaves free the leading-order non-normalizable term. On the other hand, the worldsheet metric receives a curvature correction, unless we impose $c_1 =0$ or $c_2 = 0$. Interestingly, this choices are correlated with an IR boundary condition of selecting a purely ingoing or a purely outgoing mode at the event horizon. Thus, the fluctuations can be viewed as turning on a source at the boundary, which corresponds to a diffeomorphism on the worldsheet. Thus, for all physical aspects in which the black hole is a classical object and one selects the ingoing boundary condition in the Lorentzian patch, the fluctuations can be viewed as worldsheet diffeomorphisms. Note that, the stochastic physics of the string end point, which is captured in \cite{deBoer:2008gu, Son:2009vu} by also implementing Hawking radiation of the event horizon, falls outside this class.\footnote{On an aside, given the explicit solutions, we can calculate the on-shell action of the fluctuation as well as the classical string profile and estimate whether the quadratic fluctuations remain small compared to the classical profile. In the $r\to \infty$ limit, the on-shell action receives sub-leading contribution from the fluctuations. Near $r \to r_{\rm H}$, however, the fluctuations can dominate over the classical profile. Similar to the discussion in the previous section, based on the stress tensor estimation, the fluctuations can be made sub-leading by inventing a fictitious IR cut-off\cite{deBoer:2008gu}: $\delta = r - r_{\rm H}$, which is conveniently small but non-vanishing. This cut-off is sensitive to the spectrum ({\it i.e.}~whether $\omega$ are the normal modes or the quasi-normal modes) of the system and therefore on the IR boundary condition on the fluctuations.}

%
%


\section{Branes in AdS}

In this section we will consider a wide range of examples, consisting of D-brane embedding, which, on the worldvolume description, also exhibits the saturation of the chaos bound.

\subsection{D$1$-Brane Probe: The Classical Profile}

In this section we will consider a more general class of embedding hypersurfaces, in which a similar physics will emerge. The natural candidates of such hypersurfaces are D-branes, of various dimensions. Here, we will explicitly discuss a few examples which are analytically tractable. However, the general result is expected to hold true for sufficiently generic systems. The first simplest example is to consider a rotating D$1$-brane placed in an AdS$_{d+1} \times S^{9-d}$ background, previously considered in, {\it e.g.}~\cite{Das:2010yw}. The background can be explicitly given by the following patches
\begin{eqnarray}
&& ds_{\rm Poincare}^2 = \frac{dr^2}{r^2} + r^2 \left( \eta_{\mu\nu} dx^\mu dx^\nu \right) + \left( d\theta^2 + \sin^2\theta d\phi^2 + \cos^2\theta d\Omega_{7-d}^2\right) \ , \label{AdSPoin} \\
&& ds_{\rm global}^2 = \frac{dr^2}{f(r)} - f(r) dt^2 + r^2 \sum_{i=1}^{d-1} \left(dx^i\right)^2 + \left( d\theta^2 + \sin^2\theta d\phi^2 + \cos^2\theta d\Omega_{7-d}^2\right) \ , \nonumber\\
&& f(r) = 1 + r^2 \ .
\end{eqnarray}
where $\mu, \nu = 0 , \ldots (d-1)$; $d\Omega_{7-d}^2$ is simply the metric on an $S^{7-d}$ of unit radius. For convenience, we have also set the AdS curvature scale to unity. The familiar background of AdS$_5 \times S^5$ is obtained by simply setting $d=4$. The Poincar\'{e} patch certainly covers only a part of the global section of AdS.

Let us review the embedding discussed in \cite{Das:2010yw}. We choose the gauge: $\xi^0 = t$, $\xi ^1 = r$, where $\xi^a$ ($a = 0 , 1$) represents the D$1$-brane worldvolume coordinates. To further sharpen the discussion, let us consider AdS$_5 \times S^5$ background, which is generated by a stack of D$3$-branes. The above embedding above can be represented by
\begin{table}[h!]
	\begin{center}
  			    \begin{tabular}{l|c|c|c|c|c|c|c|c|c|c|r}
			\textbf{} & $r$ & $t$ & $x^1$  & $ x^2 $ & $ x^3 $ & $ \theta $ & $ \phi $ & $ \Omega_3 $\\ 
		
			\hline
			D3 & $\times$ & $\checkmark$ & $\checkmark$ & $\checkmark$ & $\checkmark$ & $\times$ & $\times$ & $\times$ \\ 
			D1 & $\checkmark$ & $\checkmark$ & $\times $ & $\times $ & $\times$ & $\times$ & $\times$ & $\times$ \\ 
			
		\end{tabular}
	\end{center}
\end{table}
with an embedding function $\phi = \phi \left(t, r \right)$. We begin with the Poincar\'{e} patch embedding. The dynamics is governed by
\begin{eqnarray}
S_{\rm D1} = - T_{\rm D1} \int dt dr \sqrt{\frac{\sin ^2\theta  \left(r^4 \phi ^{(0,1)}(t,r)^2-\phi ^{(1,0)}(t,r)^2\right)}{r^2}+1} = - T_{\rm D1} \int dt dr \L_{\rm D1} \ , \nonumber\\ \label{D1dbi}
\end{eqnarray}
which yields the following equations of motion
\begin{eqnarray}
\partial_t \left( \frac{\sin^2\theta \phi ^{(1,0)}(t,r)}{ \L_{\rm D1} } \right)  - \partial_r \left( \frac{r^2 \sin ^2 \theta \phi ^{(0,1)}(t,r)}{ \L_{\rm D1} } \right) = 0 \ ,
\end{eqnarray}
which has the following simple solution
\begin{eqnarray}
&& \phi(t,r) = \omega t + \varphi(r) \ , \quad \theta = \frac{\pi}{2} \ , \\
&& {\rm with} \quad \varphi'(r) = \pm \frac{\sqrt{\omega ^2-r^2}}{\sqrt{r^4-c_1 r^6}} \ . \label{phip}
\end{eqnarray}
Here $c_1$ is a hitherto undetermined constant. An explicit solution for $\varphi(r)$ can be obtained in terms of complete elliptic integral, denoted by $E$, as follows:
\begin{eqnarray}
\varphi(r) = \pm \frac{\left(\sqrt{1-c_1 r^2} \left(r^2-\omega ^2\right)-r \omega  \sqrt{1-\frac{r^2}{\omega ^2}} E\left(\sin ^{-1}\left(\frac{r}{\omega }\right)|\omega ^2 c_1\right)\right)}{r \sqrt{\omega ^2-r^2}}+c_2 \ .
\end{eqnarray}
In the above $c_1$ and $c_2$ are two arbitrary constants.

Now, the constant $c_1$ can be fixed by simply demanding that the solution in (\ref{phip}) is real-valued. This enforces the numerator and denominator in (\ref{phip}) to change sign at the same location and thus fixes $c_1=1/ \omega^2 $. With this, the solution for $\varphi(r)$ becomes:
\begin{eqnarray}
\varphi(r) = c_2 \mp \frac{ \omega}{r} \ . 
\end{eqnarray}
It is straightforward to calculate the induced metric on the D$1$-brane worldvolume, which is given by
\begin{eqnarray}
ds^2 = - \left( r^2 - \omega^2 \right) d\tau^2 + \frac{dr^2}{r^2 - \omega^2 } \ , \quad {\rm with} \quad d\tau = dt + \frac{dr}{r^2} + \frac{1}{\omega^2 - r^2 } \ . \label{metd1}
\end{eqnarray}
Note that, (\ref{metd1}) is precisely the worldsheet metric in (\ref{ads2fg}), with a corresponding Hawking temperature:
\begin{eqnarray}
T_{\rm H} = \frac{\omega}{2\pi} \ .
\end{eqnarray}

Before discussing the fluctuations, let us review the brane embedding in the global section of AdS. The corresponding action is similar to (\ref{D1dbi}), with a Lagrangian density:
\begin{eqnarray}
\L_{\rm D1} = \sqrt{f(r) \sin ^2 \theta \phi ^{(0,1)}(t,r)^2 - \frac{\sin ^2 \theta  \phi ^{(1,0)}(t,r)^2}{f(r)} + 1} \ .
\end{eqnarray}
The corresponding equation of motion is given by
\begin{eqnarray}
\partial_t \left( \frac{\sin ^2 \theta \phi ^{(1,0)}(t,r)}{\L_{\rm D1}} \right)  + \partial_r \left( -\frac{f(r) \sin ^2 \theta \phi ^{(0,1)}(t,r)}{\L_{\rm D1}} \right)  = 0 \ .
\end{eqnarray}
The solution is similar to the Poincar\'{e} section solution, namely:
\begin{eqnarray}
&& \phi(t,r) = \omega t + \varphi(r) \ , \quad \theta = \frac{\pi}{2} \ , \\
&& \varphi'(r) = \frac{\sqrt{f(r)-\omega ^2}}{\sqrt{c_1 f(r)^3-f(r)^2}}  \quad \implies \quad c_1 = \frac{1}{\omega^2} \ .
\end{eqnarray}
With the constant $c_1$ fixed, the explicit solution for the embedding function is given by
\begin{eqnarray}
\phi(t, r) = \omega t \pm \omega \tan ^{-1}(r) \ . 
\end{eqnarray}
The corresponding induced metric on the worldvolume is the global section of an AdS$_2$, given by
\begin{eqnarray}
&& ds^2 = - \left( f(r) - \omega^2 \right) d\tau^2 + \frac{dr^2}{f(r) - \omega^2 } \ , \\
&& {\rm with} \quad d\tau = dt +  \frac{\omega^2}{f(r) \left( f(r) - \omega^2 \right) } dr \ , \label{metd2}
\end{eqnarray}
with a corresponding Hawking temperature:
\begin{eqnarray}
T_{\rm H} = \frac{\sqrt{\omega^2 -1}}{2\pi} \ .
\end{eqnarray}

A subsequent fluctuation analyses can be carried out like the fundamental string. The resulting four-point OTOC can be computed by evaluating the quartic term in fluctuations, and subsequently evaluating the quartic interaction, on-shell. This analysis is, qualitatively, similar to \cite{deBoer:2017xdk} and at the end, one recovers a maximal Lyapunov exponent:
\begin{eqnarray}
\lambda_{\rm L} = 2 \pi T_{\rm H} \ ,
\end{eqnarray}
where the corresponding Hawking temperature is given above. Furthermore, it is also straightforward to extract an effective Schwarzian action from the D$1$-brane theory, by embedding the D$1$-brane in an AdS$_3$ and by projecting the large diffeos of the AdS$_3$ on the brane worldvolume, similar to what we have done for the fundamental string. The presence of this Schwarzian effective action and its' coupling with other heavier modes of the D-brane fluctuations ensures maximal chaos on the brane worldvolume.

There is, however, a clear difference compared to the fundamental string. In determining the scrambling time, tension of the underlying string or brane degree of freedom directly enters. For the D$1$-brane, the corresponding scrambling time can be estimated from the following:
\begin{eqnarray}
\cO \left( T_{{\rm D}1}^{-1} e^{\lambda_{\rm L} t_{\rm sc}}\right) = \cO(1) \quad \implies \quad t_{\rm sc} \sim \frac{1}{2\pi T_{\rm H}} \log \left| \frac{N_c}{\sqrt{\lambda}}\right| \ ,
\end{eqnarray}
which is different from the scrambling time on a string worldsheet\cite{deBoer:2017xdk}: $t_{\rm sc} \sim \beta \log \sqrt{\lambda}$, where $\lambda$ is the 't Hooft coupling. Based on these generic argument, we can already estimate the scrambling time on a D$p$-brane worldvolume:
\begin{eqnarray}
t_{\rm sc} \sim \frac{1}{2\pi T_{\rm H}} \log \left| \frac{N_c}{\lambda^{\frac{3-p}{4}}}\right| \ . \label{tscgen}
\end{eqnarray}
In general, therefore, the scrambling time may be further suppressed compared to the string worldsheet, but there is always a parametric separation between the dissipation time and the scrambling time, and correspondingly the worldvolume event horizon remains the fastest scramblers for the corresponding degrees of freedom\cite{Sekino:2008he, Lashkari:2011yi}.

\subsection{D$5$-Brane Probe: The Classical Embedding}

Let us now consider a different example of a defect conformal field theory. The particular example consists of studying the dynamics of a probe D$5$-brane, as a defect degree of freedom, in the background of $N_c$ D$3$-branes. This particular defect conformal field theory was constructed and analyzed in \cite{DeWolfe:2001pq}, wherein the detailed matching of the geometric observables with the CFT observables was also identified. In brief, this describes an $\cN=4$ SYM theory in ${\mathbb R}^{3,1}$ coupled to a fundamental hypermultiplet in an ${\mathbb R}^{2,1} \subset {\mathbb R}^{3,1}$ defect. Gravitationally, this is represented by and AdS$_5 \times S^5$ geometry bisected by an AdS$_4 \times S^2$ geometry. In what follows, there is little dependence on the compact $S^5$. As long as the probe D$5$-brane can be meaningfully embedded with an AdS$_4 \subset$AdS$_5$ worldvolume, all our conclusions remain valid.

Evidently, in the dual field theory there are two distinct classes of gauge-invariant observables: the ones living in ${\mathbb R}^{3,1}$ and constructed out of the adjoint degrees of freedom; and the ones living in ${\mathbb R}^{2,1} \subset {\mathbb R}^{3,1}$ defect and constructed out of the fundamental degrees of freedom. Let us denote these two classes of observables by $\cO_4$ and $\cO_3$. In general, therefore, we can define three classes of $n$-point correlation functions, as follows:
\begin{eqnarray}
&& \C^{(1)}_n = \left \langle \cO^{(1)}_4 \ldots \cO^{(n)}_4 \right \rangle  \ , \label{ncorradj} \\
&& \C^{(2)}_n = \left \langle \cO^{(1)}_3 \ldots \cO^{(n)}_3 \right \rangle  \ , \label{ncorrfund} \\
&& \C^{(3)}_{p, q} = \left \langle \cO^{(1)}_4 \ldots \cO^{(p)}_4 \cO^{(1)}_3 \ldots \cO^{(q)}_3 \right \rangle  \ , \quad {\rm with} \quad p+q = n \ . \label{ncorradjfund}
\end{eqnarray}
In our discussion, we will explicitly calculate the $4$-point OTOC in class (\ref{ncorrfund}). We will consider purely AdS$_5\times S^5$ background, and therefore the $n$-point correlation function of class (\ref{ncorradj}) will have no thermal imprint. Furthermore, the $n$-point correlation of class (\ref{ncorradjfund}) are suppressed in our approximation: $N_f \ll N_c$. Here $N_f$ and $N_c$ are the number of D$5$ and D$3$-branes, respectively.

The geometry of AdS$_5\times S^5$ is described by the following patch:
\begin{eqnarray}
&& ds^2 = \frac{r^2}{R^2} \left(-f(r) dt^2 + d\vec{x}^2 \right) +\frac{R^2}{r^2 f(r)} dr^2 + R^2 d\Omega_5^2  \ , \\
&& d\Omega_5^2 = d\psi^2 + \cos^2\psi \left(d\theta^2 + \sin^2\theta d\phi^2 \right) + \sin^2\chi \left(d\chi^2 + \sin^2\chi ~ d\zeta^2\right) \ , \\
&& f(r) = 1 - \left(\frac{r_{\rm H}}{r}\right)^4 \ . 
\end{eqnarray}
In the above, $\{t, \vec{x}\}$ represents the ${\mathbb R}^{1,3}$, where the $\cN=4$ SYM is defined. 
For general purposes, we have kept a non-vanishing event horizon, denoted by $r_{\rm H}$ and a corresponding temperature: $T = r_{\rm H}/ (2\pi R^2)$. The relative orientation of the D$3$ and the D$5$ branes can be represented in the table below:
\begin{table}[h!]
	\begin{center}
  			    \begin{tabular}{l|c|c|c|c|c|c|c|c|c|c|r}
			\textbf{} & $r$ & $t$ & $x^1$  & $ x^2 $ & $ x^3 $ & $ \psi $ & $ \theta $ & $ \phi $ & $ \chi $ & $ \zeta $\\ 
		
			\hline
			D3 & $\times$ & $\checkmark$ & $\checkmark$ & $\checkmark$ & $\checkmark$ & $\times$ & $\times$ & $\times$ & $\times$ & $\times$\\ 
			D5 & $\checkmark$ & $\checkmark$ & $\checkmark$ & $\checkmark$ & $\times$ & $\times$ & $\checkmark$ & $\checkmark$ & $\times$ & $\times$\\ 
			
		\end{tabular}
	\end{center}
\end{table}
To proceed further, we need to define embedding functions. We choose the following embedding function data:
\begin{eqnarray}
\psi = \psi(r) \ , \quad  x^3 = 0 \ , \quad \chi = {\rm constant} \ , \quad  \zeta =  {\rm constant} \ .
\end{eqnarray}
The $x^3 = 0$ slice picks out the ${\mathbb R}^{2,1} \subset {\mathbb R}^{3,1}$ defect and the two angular coordinates simply imply that the D$5$-brane is a point on the corresponding $S^2$. The dynamics is governed by a DBI-action, similar to the D$1$ probe situation:
\begin{eqnarray}
S_{{\rm D}5} &=& - T_{{\rm D}5} \int d^6\xi \sqrt{- {\rm det} \left( P[G] + \left( 2 \pi \alpha' \right) F \right)} + g_s T_{{\rm D}5} \int P \left[C_4 \right]\wedge \left( 2 \pi \alpha' \right) F \ , \nonumber\\
& = & - T_{{\rm D}5} \int d^6\xi \L_{\rm D5} \ .
\end{eqnarray}
where $F$ is the worldvolume $U(1)$ gauge field, $C_4$ is the RR-potential sourced by the background D$3$ branes. In general the Lagangian density is a functional of the embedding function $\psi(r)$, and the $U(1)$ gauge field, provided we choose to excite the latter. To induce an event horizon in the probe sector, we will excite an appropriate component of the gauge field, following a similar idea in {\it e.g.}~\cite{Karch:2007pd, Erdmenger:2007bn, Albash:2007bq, Alam:2012fw, Sonner:2012if}. Let us choose:
\begin{equation}
A = \left(- E t - a_x(r) \right) dx^1 \ , \quad \implies \quad F = - E ~dt\wedge dx^1 - a_x'(r) ~dr\wedge dx^1 \ .
\end{equation}
It is straightforward to check that: $P[C_4]\wedge F = 0$ and therefore:
\begin{eqnarray}
\L_{\rm D5} = \L_{\rm D5}\left[ \psi(r), \psi' (r), a_x'(r)\right] \ .	
\end{eqnarray}
The resulting Euler-Lagrange equations of motion are given by
\begin{eqnarray}
\frac{d}{dr} \left[ \frac{\partial \L_{\rm D5}}{\partial \psi'(r)} \right]  - \frac{\partial \L_{\rm D5}}{\partial \psi(r)} = 0 \ ,  \quad \frac{d}{dr} \left[ \frac{\partial \L_{\rm D5}}{\partial a_x'(r)} \right] = 0 \ . \label{EuLag}
\end{eqnarray}
The general solution of (\ref{EuLag}) represents a class of functions $\psi(r)$ characterized by its asymptotic boundary values. For convenience, we will focus on the following solution:
\begin{eqnarray}
\psi(r) &=& 0 \ , \\
a_x'(r) &=& \frac{D r^2}{\left(r^4 - r_{\rm H}^4\right)} \sqrt{\frac{\left(r^4 - (r_{\rm H}^4 + E^2R^4)\right)}{\left(r^4 - (r_{\rm H}^4+D^2)\right)}} \ , 
\end{eqnarray}	
The reality condition of the function $a_x'(r)$ now imposes $D = E R^2$. Note that, this is independent of the value of $r_{\rm H}$. 

The physical picture is simple to understand. The scalar function $\psi(r)$, asymptotically has the information of source and the vev in the corresponding dual field theory. In fact, asymptotically, the solution of the two bulk fields $\psi(r)$ and $a_x'(r)$ can be obtained as:
\begin{eqnarray}
&& \psi(r) = c_1 r^{-1} + c_2 r^{-2} + \ldots \ , \\
&& a_x'(r) = c_3 r^{-2} + \ldots \ .
\end{eqnarray}
In general, the non-normalizable mode in $\psi(r)$ correspond to the mass of the probe sector fundamental degree of freedom and the normalizable mode corresponds to the corresponding vev, which correspond to the constants $c_1$ and $c_2$, respectively. A similar identification holds true for $c_3$, which corresponds to a current in the boundary theory. It can be easily checked that $c_3$ is proportional to the constant $D$.

The open string degrees of freedom correspond to bound states of quark-like matter\cite{Karch:2002sh}. These are typically mesonic bound states. With our description above, there is no free fundamental matter. This can be incorporated by further exciting an appropriate gauge field on the D$5$-brane worldvolume, which we do not consider here. Thus, in the presence of a sufficiently large electric field, there will be pair creation resulting in a  matter current. Clearly, this critical electric field depends on the mass of the pair created matter. In brief, the mesonic and the conducting phases are connected by a first order phase transition, similar to the ones studied in \cite{Erdmenger:2007bn, Albash:2007bq, Alam:2012fw}. In the subsequent discussion here, we focus on the massless case. This is motivated by the analytical control in this corner. In principle, all subsequent calculations in the probe fluctuation sector can be carried out for a general massive case, with little qualitative difference.

\subsection{D$5$-Brane Probe: Fluctuations}

Similar to the previous cases, we will now discuss the physics of the fluctuation fields on the D-brane worldvolume. The fluctuations can be divided into three categories: (i) scalar, (ii) vector and (iii) spinors. We will explicitly consider only the first two kind and further truncate them to a sector which is analytically tractable. Schematically, let us denote the worldvolume fields as follows:
\begin{eqnarray}
X^I &=& X^I_{\rm cl} + \delta X^I \left( \xi^a \right ) \ , \\
A_b &=& A_b^{\rm cl} + \delta A_b \left(\xi^a \right) \ .
\end{eqnarray} 
In the above, $I = 6, \ldots, 9$ (transverse to the D$5$-brane) and $b = 0, \ldots 4$ (parallel to the D$5$-brane). Collectively, $\xi^a$ denotes the D$5$-brane worldvolume directions.

These scalar fluctuations (along the transverse directions of the D$5$-brane) lead to fluctuations of the induced metric:
\begin{eqnarray}
\delta \gamma_{ab} & = & 2 \left(\partial_a X^{I}_{\rm cl}~ \partial_b \delta X^{J}~ G_{IJ}\right) + \left( \partial_a \delta X^I\right) \left(\partial_b \delta X^J \right) G_{IJ} \nonumber \\
	&=& \left( \partial_a \delta X^I \right) \left( \partial_b \delta X^J \right) G_{IJ} \ ,
\end{eqnarray}
where, the first term drops out because, for our case $X^I_{\rm cl}$ is either zero or constant. Here $G_{IJ}$ are the background metric components along the directions transverse to the D$5$ brane. Similarly, the vector fluctuations lead to fluctuation of the field strength:
\begin{equation}
\delta F_{ab} = \partial_a \delta A_b - \partial_b \delta A_a \ .
\end{equation}
In what follows we will absorb the factor of $\left( 2 \pi\alpha'\right)$ inside the worldvolume field strength, for convenience. Note that $\delta\gamma_{ab}$ is quadratic in fluctuations whereas $\delta F_{ab}$ is linear. To find the dynamics of the fluctuations we look to expand the DBI action upto quadratic order in fluctuations.

This quadratic fluctuation piece can be calculated as below:
\begin{equation}
\begin{aligned}
S_{{\rm D}5} = & - T_{{\rm D}5} \int d^6 \xi \sqrt{- {\rm det} \left( \cM + \delta \cM \right)} \ , \quad  \cM = P\left[G \right] + F \\
 = & - T_{{\rm D}5} \int d^6 \xi \sqrt{-{\rm det} \cM}~ \sqrt{{\rm det} \left(1 + \X \right)} \ , \quad  \X = \cM^{-1}\delta \cM \\
= & - T_{{\rm D}5} \int d^6 \xi \sqrt{-{\rm det} \cM}~ \sqrt{{\rm exp} \left[ {\rm Tr} \log \X \right]} \\
\approx & - T_{{\rm D}5} \int d^6 \xi \sqrt{-{\rm det} \cM} ~ \left(1 + \frac{1}{2} {\rm Tr} \X + \frac{1}{8} ({\rm Tr} \X)^2  -\frac{1}{4} {\rm Tr} \X^2\right) + \cO \left( \X^3 \right) \ . \label{flucD5}
\end{aligned}
\end{equation} 
Now since  $\delta\gamma_{ab}$ is quadratic while $\delta F_{ab}$ is linear in fluctuation we shall  check what contributions we have from the above expansion at every order.

\subsection{At First Order}

At first order, only ${\rm Tr}\X$ will contribute and this gives a contribution in the vector sector. Note that:
\begin{eqnarray}
\nonumber {\rm Tr}\X &=& \cM^{ab} \delta \cM_{ba} =  \S^{ab} \delta \gamma_{ab} + \mathcal{A}^{ab}\delta F_{ab} \ .
\end{eqnarray}
Here $\S$ and $\A$ are completely symmetric and completely anti-symmetric matrices, respectively. So, clearly the second term contributes at the first order. However, on-shell, this vanishes:
\begin{eqnarray}
\nonumber	S_{{\rm D}5}^{(1)} &\approx& - T_{{\rm D}5} \int d^6 \xi \sqrt{-{\rm det} \cM} \, \mathcal{A}^{ab}\delta F_{ab} + \cO\left( \delta F^4 \right) \\
\nonumber &=& -  2 T_{{\rm D}5} \int d^6 \xi \sqrt{-{\rm det} \cM}  \mathcal{A}^{ab} \partial_a \delta A_{b} \\
 &=&   2 T_{{\rm D}5} \int d^6 \xi  ~ \partial_a\left( \sqrt{-{\rm det} \cM} \mathcal{A}^{ab}\right) \delta A_b + {\rm boundary-terms} \ .
\end{eqnarray}	
Thus, $S_{{\rm D}5}^{(1)}$ vanishes on-shell.

\subsection{At Second Order}

Let us again check term by term. ${\rm Tr}\X$ now yields a contribution, which comes from the term $\S^{ab}\delta \gamma_{ab}$. Now
\begin{equation}
\left( {\rm Tr}\X \right)^2 =  {\rm Tr} \left(\S.\delta\gamma\right)  {\rm Tr} \left(\S. \delta\gamma\right) + 2~ {\rm Tr} \left(\S.\delta\gamma\right) {\rm Tr} \left(\mathcal{A}. \delta F \right) + {\rm Tr} \left(\mathcal{A} .\delta F \right) {\rm Tr} \left(\mathcal{A} .\delta F \right) \ ,
\end{equation}
where the `dot' product is simply defined as the matrix product. The first term is quartic in fluctuations whereas the second term is third order in fluctuations. At quadratic order, only the last term contributes. At this order, we have another contribution coming from the last term in (\ref{flucD5}), given by
\begin{equation}
\begin{aligned}
\nonumber {\rm Tr}\X^2 = {} & \{ {\rm Tr} \left( \S.\delta \gamma.\S \delta \gamma \right) + {\rm Tr} \left(\mathcal{A}.\delta\gamma.\mathcal{A}.\delta\gamma \right)\} + \{  {\rm Tr} \left( \S. \delta \gamma.\mathcal{A}.\delta F \right) + 2~ {\rm Tr} \left( \S.\delta F.\mathcal{A}.\delta\gamma \right) \\
& {\rm Tr} \left(\mathcal{A}.\delta F . \S .\delta\gamma \right)\} + \{{\rm Tr} \left( \S .\delta F. \S . \delta F \right) + {\rm Tr} \left(\mathcal{A}.\delta F. \mathcal{A} . \delta F \right) \} 
\end{aligned}
\end{equation}
Here first two terms are quartic in fluctuations and next three terms are third order in fluctuations. So only the last two terms will contribute.

So finally, we get the following quadratic action:
\begin{equation}
\begin{aligned}
S_{{\rm D}5}^{(2)} = &  - T_{{\rm D}5} \int d^6 \xi \sqrt{- {\rm det} \cM} \left [ \S^{ab} \delta \gamma_{ab} + \right.\nonumber\\
&\left.  \frac{1}{8} {\rm Tr} \left(\mathcal{A} .\delta F \right) {\rm Tr} \left(\mathcal{A} .\delta F \right) - \frac{1}{4} \left( {\rm Tr} \left( \S .\delta F. \S. \delta F \right) + {\rm Tr} \left(\mathcal{A}.\delta F. \mathcal{A} . \delta F \right) \right ) \right ] \\
=&  - T_{{\rm D}5} \int d^6 \xi \sqrt{- {\rm det} \cM} \left( \S^{ab} \delta\gamma_{ab}   -\frac{1}{4} {\rm Tr} \left( \S .\delta F. \S. \delta F \right ) \right) \ ,
\end{aligned}
\end{equation}
where the second equality follows from the fact that 
\begin{eqnarray}
\frac{1}{8} {\rm Tr} \left(\mathcal{A} .\delta F \right) {\rm Tr} \left(\mathcal{A} .\delta F \right) = \frac{1}{4} {\rm Tr} \left(\mathcal{A}.\delta F. \mathcal{A} . \delta F \right) \ . 
\end{eqnarray}
Note that, the above equality holds true even off-shell. Thus, the scalar and the vector fluctuations decouple at the second order. From now on, we set the scalar fluctuations to zero and focus only on the vector degrees of freedom. Therefore, the corresponding Lagrangian is given by
\begin{equation}
S_{{\rm vector}}^{(2)} = - \frac{1}{4} T_{{\rm D}5} \int d^6 \xi \sqrt{- {\rm det} \cM} \, {\rm Tr} \left( \S .\delta F. \S. \delta F \right ) \ . \label{vecfluc}
\end{equation}
The corresponding equation of motion are given by
\begin{equation}
\partial_a \left( \sqrt{- {\rm det} \cM}~ \S^{ab}~\delta F_{bc} ~ \S^{cd}\right) = 0 \ . \label{genveceqn}
\end{equation}
We will solve these equations in the Poincar\'{e} patch as well as in the corresponding Kruskal coordinate.

\subsubsection{Specifying Data}

Recall that: 
\begin{equation}
\nonumber	\cM_{ab} = \gamma_{ab} + F_{ab} \ , 
\end{equation}
where $\gamma = P[G]$. Therefore:
\begin{equation}
\nonumber	\cM^{ab} = \S^{ab} + \mathcal{A}^{ab} \ ,
\end{equation}
where
\begin{eqnarray}
\S^{ab} & = & \left[(\gamma+F)^{-1} .~ \gamma . ~(\gamma - F)^{-1} \right]^{ab} \ , \label{osmup} \\
\mathcal{A}^{ab} & = & - \left[(\gamma+F)^{-1} .~ F . ~(\gamma - F)^{-1} \right]^{ab} \ .
\end{eqnarray}
The so-called open string metric, denoted by $\S_{ab}$, is the inverse of (\ref{osmup}) and is given  by
\begin{equation}
\S_{ab} = \gamma_{ab} - \left[F.~ \gamma^{-1}.~ F\right]_{ab} \ .
\end{equation}
Given the background geometry, the embedding and the worldvolume gauge field, we can compute the components of the osm. In this case, it turns out to be off-diagonal in the $\{r, t\}$-plane. However, under the coordinate transformation $ t \rightarrow \tau  = t + f(r) ,~ r \rightarrow R$ and then relabelling $(\tau , R)$ as $(t,r)$, the osm can be diagonalized in the following explicit form:
\begin{equation}
ds^2_{\rm osm} = - \frac{r^4- r_*^4}{R^2 r^2} dt^2 + \frac{R^2 r^2}{r^4- r_*^4} dr^2 + \frac{r^2}{R^2} \left(dx_1^2 + dx_2^2\right) +  R^2 \left(d\theta^2 + \sin^2\theta d\phi^2 \right) \ .
\end{equation}
Here $r_*^2 = D^2 + r_{\rm H}^4$ along with $D^2 = E^2 R^4$. We will consider the diagonal metric. On the other hand, the components of $\mathcal{A}^{ab}$ are given by 
\begin{equation}
		\mathcal{A}^{ab} = \left(
		\begin{array}{cccccc}
		0 & 0 & \frac{D}{r^2} & 0 & 0 & 0 \\
		0 & 0 & \frac{DR^2}{r_{\rm H}^4 - r^4} & 0 & 0 & 0 \\
		-\frac{D}{r^2} & -\frac{D R^2}{r_{\rm H}^4 - r^4} & 0 & 0 & 0 & 0 \\
		0 & 0 & 0 & 0 & 0 & 0 \\
		0 & 0 & 0 & 0 & 0 & 0 \\
		0 & 0 & 0 & 0 & 0 & 0 \\
		\end{array}
		\right)
\end{equation}
Here, we have used the explicit solution for the gauge fiels: $a_x'(r) = \frac{D r^2}{r^4-r_{\rm H}^4}$. Now, we consider the following modes of the fluctuation:
\begin{equation}
\delta A = \delta A_t (t,r)~ dt + \delta A_{||}(t,r)~ dx^1 + \delta A_{\perp}(t,r) ~dx^2 \ . \label{gaugefluc}
\end{equation}
Here, the symbols $||$ and $\perp$ corresponds to gauge modes parallel and perpendicular to the electric field, respectively. With all these ingredients we look to solve the equations of motion in both $(t,r)$.

\subsubsection{Solution in $(r,t)$ Coordinates}

The equation of motion in covariant form reads:
\begin{equation}
\partial_a \left(\sqrt{- {\rm det} \cM} \, \S^{ab}~\delta F_{bc}~ \S^{cd}\right) = 0 \ ,
\end{equation}
where $\delta F_{ab}$ is obtained from the gauge fluctuations in (\ref{gaugefluc}). Written in components, these take the following form:
\begin{equation}
\underline{\textbf{$r$  equation}}: \quad  \partial_t \left (r^2~ \partial_r \delta A_t \right) = 0 \ . 
\end{equation}
\begin{equation}
\underline{\textbf{$t$  equation}}: \quad \partial_r \left (r^2~ \partial_r \delta A_t \right) = 0 \ . 
\end{equation}
On solving these two we get:
\begin{equation}
\delta A_t (r,t) = c_1 - \frac{c_2}{r} \ . 
\end{equation}
This mode will not have any dynamics and we shall set this mode to be identically zero ($c_1 = 0, ~ c_2 = 0$).

Let us now look at the following:
\begin{equation}
\underline{\textbf{$x^1$  equation}}:	\quad \partial_t \left (r^2 \S^{tt} \left(\partial_t \delta A_{\parallel}\right) \S^{x_1 x_1} \right) + \partial_r \left (r^2 \S^{rr} \left(\partial_r \delta A_{\parallel} \right)  \S^{x_1 x_1} \right) = 0 \ .
\end{equation}
We look for a solution $\delta A_{\parallel} (r,t ) = f(r) e^{i \omega t} $. Plugging this in, we get:
\begin{equation}
r \left(r^4-r_*^4\right)^2 f''(r) + 2 \left(r^8 - r_*^8\right) f'(r)+ R^4 r^5 \omega ^2 f(r) = 0 \ . 
\end{equation}
This equation has the solution:
\begin{equation}
\delta A_{\parallel}(t,r) = \left(C_1~ e^{i \omega  \tilde{r} }+ C_2 ~e^{- i \omega \tilde{r} } \right) e^{i \omega  t} \hspace{5mm}  \label{solvecfluc}
\end{equation} 
where $\tilde{r}$ is the tortoise coordinate we will shortly describe. Finally, we look at the $x^2$-equation: Because of the isotropy in the $\{x^1- x^2\}$-plane, $x^2$ equation is also identically same as  the previous equation with the same solution. Thus, 
\begin{eqnarray}
\delta A_{\perp} (t,r) = \delta A_{\parallel}(t,r)  \ . \left(C_1~ e^{i \omega  \tilde{r} }+ C_2 ~e^{- i \omega \tilde{r} } \right) e^{i \omega  t}  \ . \hspace{5mm}  \label{solvecflucperp}
\end{eqnarray}
We will now discuss the Kruskal patch analysis.

\subsubsection{Kruskal extension and Solution in Kruskal Patch}

To find the explicit Kruskal patch, it is sufficient to look at the $\{t, r \}$-plane of the osm geometry. This is given by
\begin{eqnarray}
\nonumber 	ds^2 & = & - \frac{r^4 - r_{\rm E}^4}{R^2 r^2} dt^2 + \frac{R^2 r^2}{r^4 - r_{\rm E}^4} dr^2 + \ldots \\
& = & \frac{r^4 - r_{\rm E}^4}{R^2 r^2} \left( - dt^2 + dr_{*}^2 \right) \ ,
\end{eqnarray}
where the tortoise coordinate is given by 
\begin{equation}
r_{*} = \int dr_{\star} = \int \frac{R^2 r^2}{r^4 - r_{\rm E}^4} dr = \frac{R^2}{4 r_{\rm E}} \left[2 \tan^{-1}\left(\frac{r}{r_{\rm E}}\right) + \log \left(\frac{r - r_{\rm E}}{r + r_{\rm E}} \right) \right] \ . \label{tortoiser}
\end{equation}
Now, we define the Eddington-Finkelstein coordinates as:
\begin{eqnarray}
&& v = t + r_{*} \ , \\
&& u = t - r_{*} \ .
\end{eqnarray}
Subsequently, the Kruskal coordinates are: 
\begin{eqnarray}
V &=& {\rm exp} \left[ \frac{2 v r_{\rm E}}{R^2}\right] = \sqrt{\frac{r - r_{\rm E}}{ r + r_{\rm E}}}~ {\rm exp}\left[{\frac{2 t r_{\rm E}}{R^2} + \tan ^{-1}\left(\frac{r}{r_{\rm E}}\right)}  \right] \\
U &=& - {\rm exp} \left[ - \frac{2 u r_{\rm E}}{R^2}\right] = - \sqrt{\frac{r - r_{\rm E}}{r + r_{\rm E}}}~ {\rm exp} \left[{\tan ^{-1}\left(\frac{r}{r_{\rm E}}\right) - \frac{2 t r_{\rm E}}{R^2}}\right] \ . 
\end{eqnarray}
Finally, we define:
\begin{eqnarray}
T &=& \frac{1}{2}\left(V + U\right) = \sqrt{\frac{r - r_{\rm E}}{ r + r_{\rm E}}}~ \exp \left[\tan ^{-1}\left(\frac{r}{r_{\rm E}}\right) \sinh \left(\frac{2 r_{\rm E} t}{R^2}\right)\right] \ , \\
Y &=& \frac{1}{2}\left(V - U\right) = \sqrt{\frac{r - r_{\rm E}}{r + r_{\rm E}}} ~\exp \left[ \tan ^{-1}\left(\frac{r}{r_{\rm E}}\right) \cosh \left(\frac{2 r_{\rm E} t}{R^2}\right)\right]
\end{eqnarray}

Now given the above map, we need to  write the components of $\S_{ab}$, $\delta F_{ab}$ and $\mathcal{A}^{ab}$ in the $\{T,Y\}$ coordinates. The osm is given by (using starndard coordinated transformation rule)
\begin{eqnarray}
&& ds^2 = \Omega(r) \left(-dT^2 + dY^2\right) + \frac{r^2}{R^2} \left(dx_1^2 + dx_2^2\right)+  R^2 \left(d\theta^2 + \sin^2\theta d\phi^2 \right) \ , \\
&& \Omega(r) = \frac{R^2 \left(r_{\rm E} + r \right){}^2 \left(r^2 + r_{\rm E}^2\right) {\rm exp} \left[{-2 \tan ^{-1}\left(\frac{r}{r_{\rm E}} \right)} \right]} {4 r^2 r_{\rm E}^2} \ .
\end{eqnarray}
As far as the fluctuations are concerned we saw that $\delta A_{t}$ has no dynamics. On the other hand, because of the isotropy in the $\{x^1, x^2\}$-plane, the parallel and the transverse fluctuations have exactly identical solution at the quadratic order. Furthermore, all these modes are decoupled from each other. From now on, we will concentrate only on the transverse fluctuation and set  the remaining two modes to zero. This induces a drastic simplification in the subsequent analysis, since
\begin{equation}
{\rm Tr} \left(\cM^{-1} \delta \cM \right) = \mathcal{A}^{ab} \delta F_{ab}  = 0 \ ,
\end{equation}
which holds identically, even without performing the integration. Moreover, we will now write $ \delta A = \delta A_{\perp} \left(T,Y \right) dx^2 $, which is related to the Poincar\'{e} patch solution {\it via} the map: $T= T(r,t)$ and $Y= Y(r,t)$. Also, the explicit components of $\A^{ab}$ are quite complicated in the $\{Y, T\}$-coordinates so we are not presenting those.

Gathering all these, in the  Kruskal coordinates, the quadratic action (\ref{vecfluc}) now becomes:
\begin{eqnarray}
\nonumber 	S_{{\rm vector}}^{(2)} & = & \frac{1}{2} T_{{\rm D}5}  \int dT dY~ R^2   \left(\delta \dot{ A}^2 (T,Y)- \delta{ A'}^2 (T,Y)\right)   \\
& = & \frac{1}{2} \int dT dY ~\left( \delta \dot{ A}_{\rm ren}^2 (T,Y) - \delta { A'}_{\rm ren}^2 (T,Y)\right) \ . \label{resdeltaA}
\end{eqnarray}
where we have redefined the fields as $\delta A_{\rm ren} \equiv R \sqrt{T_{{\rm D}5}} ~\delta A ~ $ to canonically normalize the kinetic term. Also, $~$ $\dot{} \equiv \partial_{T}$ , $ ~ ' \equiv \partial_Y $. We can freely drop the subscript ``ren" from the redefined fluctuations. This results in the following equation of motion:
\begin{equation}
\left(\partial_T^2 - \partial_Y^2\right) \delta A = 0 \ , \label{eqnkrus}
\end{equation}
which has the solution:
\begin{equation}
\delta A = F(T+Y) + F(T - Y) \ .  \label{gensoldeltaA}
\end{equation}
Here $F$ is a completely arbitrary function, desired to be smooth on physical grounds. Note that this solution resembles (\ref{solvecfluc}), written in Kruskal coordinates. Also (\ref{eqnkrus}) follows from  the general covariant equations in (\ref{genveceqn}), by invoking the Poincar\'{e} to Kruskal map.

\subsection{At $4^{\rm th}$ Order}

To compute the $4$-point OTOC, we need to calculate the four-point interaction term in the probe sector. This can be simply done by evaluating the quartic piece in the fluctuation Lagrangian, and subsequently setting this term on-shell with respect to the quadratic action. Let us now evaluate the quartic interaction piece. We will set the scalar fluctuations to zero and also set ${\rm Tr} \X = {\rm Tr} \left( \cM^{-1} \delta \cM \right) = 0$. The last condition eliminates a cubic interaction term which is otherwise present in the fluctuation Lagrangian. Expanding the DBI action upto fourth order, we get:
\begin{eqnarray}
\nonumber 	S_{{\rm D}5}^{(4)} & = & - T_{{\rm D}5} \int d^6 \xi \sqrt{- {\rm det} \cM} \left(\frac{1}{32} \left({\rm Tr} \X^2\right)^2 - \frac{1}{8} {\rm Tr} \X^4\right) \\
 & = & \frac{N~e^{\pi /2} }{16 r_{\rm E}^2~ R^2 T_{{\rm D}5} } \int dT dR \left(\delta \dot{ A }_{\rm ren}^2-\delta A_{\rm ren}'^2\right)^2 \ . \label{deltaAquartic}
\end{eqnarray}
The last line above, we have used the rescaled $\delta A_{\rm ren}$, defined below (\ref{resdeltaA}). Also, (\ref{deltaAquartic}) has been evaluated near the horizon, {\it i.e.}~at $U \approx 0 $ and $V \approx 0$. The constant, denoted by $N$, represents an $\mathcal{O}(1)$ constant, coming out of the integral over the angular coordinates. Substituting ({\ref{gensoldeltaA}}), we can recast this quartic action as:
\begin{equation}
S_{{\rm D}5}^{(4)}  = \frac{N~e^{\pi /2} }{2 r_{\rm E}^2~ R^2 T_{{\rm D}5} } \left( \int d \zeta  F'(\zeta)^2\right)^2 \ . \label{S4th}
\end{equation} 
At this order, the contribution is schematically similar to the one obtained in {\it e.g.}~\cite{deBoer:2017xdk}.

\section{$4$-point OTO Correlation}

%
\begin{figure}[ht!]
\begin{center}
{\includegraphics[width=1\textwidth]{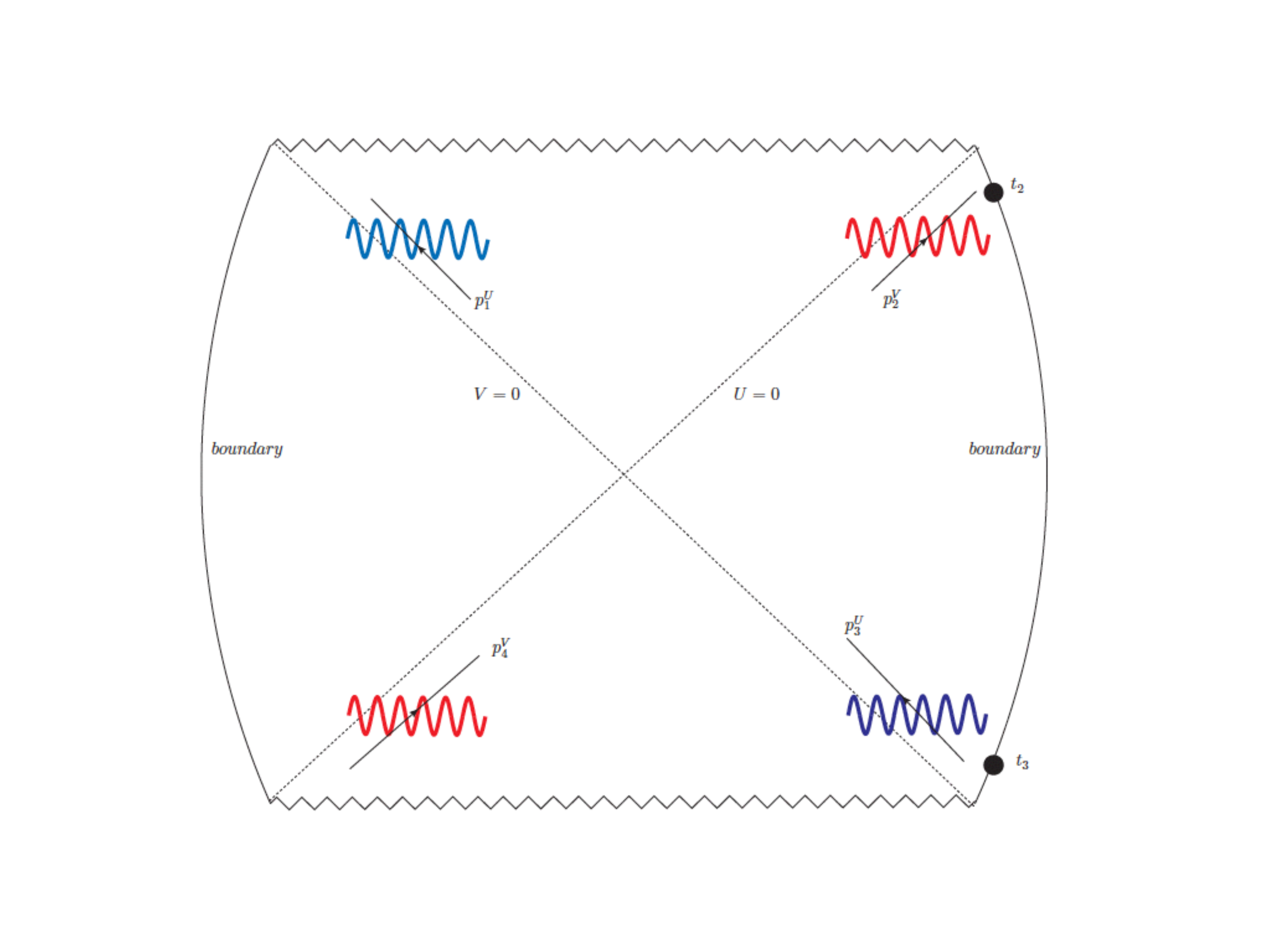}}
\caption{\small A Penrose diagram representation of the $2-2$ elastic scattering which is used to calculate the phase shift. This is identical to the discussion in \cite{Shenker:2014cwa}. The Kruskal extension was obtained in \cite{Banerjee:2016qeu}.} \label{2-2penrose}
\end{center}
\end{figure}
A schematic pictorial representation of the $2-2$ scattering process is demonstrated in figure \ref{2-2penrose}. We are interested in computing the OTOC of the form: $\langle A(t_4) A(t_3) A(t_2)  A(t_1) \rangle$,  where $A$ is the boundary operator corresponding to the field  $\delta A_{\rm ren}$, which was introduced in the previous section. We are considering a thermal expectation value at temperature 
\begin{eqnarray}
T = \beta^{-1} = \frac{r_{\rm E}}{\pi R^2} \ , 
\end{eqnarray}
which is the temperature associated with the osm horizon.

This calculation, following the treatment in \cite{Shenker:2014cwa}, can be set up as a $2-2$ scattering amplitude computation in the corresponding thermofield double state. From a geometric point of view, this eventually corresponds to the elastic scattering process between two incoming high-energy pulses into two outgoing high-energy modes. Finally, this scattering amplitude in the thermofield double state can be re-interpreted in terms of the one-sided thermal correlation function.

Thus, the ``in" state corresponds to two high energy modes, with momenta $p_3^U$ and $p_4^V$; the ``out" state corresponds to two high energy modes, with momenta $p_1^U$ and $p_2^V$, respectively. Therefore, the in and out states can be expanded in longitudinal momentum basis as follows:
\begin{eqnarray}
A(t_3)~ A(t_4) \left |\beta \right \rangle &=& \int dp_3 ^U dp_4^V ~ \psi_3 \left(p_3^U\right) \psi_4 \left(p_4^V\right) ~ \left| p_3^U p_4^V \right\rangle_{\rm in}\\
A(t_2)~ A(t_1) \left|\beta \right \rangle &=& \int dp_1 ^U dp_2^V ~ \psi_1 \left(p_3^U\right) \psi_2 \left(p_4^V\right) ~ \left| p_1^U p_2^V \right \rangle_{\rm out} \ ,
\end{eqnarray}
where the $\{\psi\}$ are the wavefunctions of the corresponding field dual to the operator. Mathematically, they are given by the Fourier transforms of the corresponding bulk-to-boundary propagators, as explained in \cite{Shenker:2014cwa}. The normalisation of the basis states are given by 
\begin{equation}
\langle p^V | q^V \rangle = \frac{4 p^V}{\pi} ~\delta\left( p^V - q^V\right) \ , \quad \langle p^U | q^U \rangle = \frac{4 p^U}{\pi} ~\delta\left( p^U - q^U \right) \ , 
\end{equation}
We are interested in high-energy $2-2$ elastic scattering. Elastic scattering implies that $p_1^U \approx p_3^U$, $p_2^V \approx p_4^V$, which is a simple consequence of momentum conservation along $U$ and $V$ directions. Furthermore, the elastic scattering implies that the ``in" state and the ``out" state differ only by a phase:
\begin{eqnarray}
\left| p_1^U p_2^V \right\rangle_{\rm out} \approx e^{i \delta(s)} \left| p_1^U p_2^V \right \rangle_{\rm in} \ ,  
\end{eqnarray}
where $s = 4 p_1^U p_2^V $ is the corresponding Mandelstam variable. Thus, finally, the OTOC is given by the overlap integral:
\begin{equation}
\left \langle A(t_4) A(t_3) A(t_2)  A(t_1)\right \rangle = \int dp_1^U dp_2^V ~ p_1^U p_2^V~ e^{i \delta (s)} ~\psi_4^{*}\left(p_2^V \right) \psi_3^{*}\left(p_1^U \right) \psi_2\left(p_2^V \right) \psi_1\left(p_1^U\right) \ ,
\end{equation}
where we have explicitly used momentum conservation conditions. One, therefore, needs to calculate the phase factor and the corresponding wave functions.

\subsection{Phase factor}

From the elastic eikonal gravity approximation, we immediately have $\delta(s) = S_{{\rm D}5}^{(4)}$, where $S_{{\rm D}5}^{(4)}$ is given by ({\ref{S4th}}). Now, the energy of the wave packets is given by 
\begin{equation}
\Delta E = 2 \int d \zeta F'(\zeta)^2 \ , \label{energy}
\end{equation}
and the corresponding Mandelstam variable is $s = (\Delta E)^2 $. There are several ways to obtain (\ref{energy}). Perhaps the simplest is to notice that the fluctuation Lagrangian in (\ref{resdeltaA}) is given by a term $\left( \delta \dot{ A}_{\rm ren}^2 - \delta { A'}_{\rm ren}^2 \right)$, which yields a Hamiltonian, given by $\left( \delta \dot{ A}_{\rm ren}^2 + \delta { A'}_{\rm ren}^2 \right)$. Taking into account the solution in (\ref{gensoldeltaA}), and appropriately accounting the Jacobian corresponding to a co-ordinate transformation, one simply arrives at (\ref{energy}). So, using all these we finally get:
\begin{equation}
\delta(s) = \frac{N~e^{\pi /2} }{8 r_{\rm E}^2~ R^2 T_{{\rm D}5}}~s = \frac{N~e^{\pi /2} }{2 r_{\rm E}^2~ R^2 T_{{\rm D}5}}~ p_1^U p_2^V \ .
\end{equation}
The phase shift, therefore, has an expected functional dependence with the Mandelstam variable $s$, while the coefficient is determined by the tension of the D-brane. Evidently, we assume that, for this high energy scattering process, the DBI-action is a valid description and one thus obtains: $\delta (s) \sim s$.

\subsection{Wave function \& Bulk-to-boundary Propagator}

Given two bulk points $\{t, r\}$ and $\{t', r'\}$, we can construct the bulk-to-bulk propagator for the corresponding vector fluctuation, making use of the two analytically known solutions of the given fluctuation equation at the quadratic order. The corresponding solutions are given in (\ref{solvecfluc}).

The bulk-to-bulk propagator, denoted by $G\left(t, r; t', r' \right)$ satisfies the following equation:
\begin{eqnarray}
\L G\left(t, r; t', r' \right) = \delta \left( t - t' \right) \delta \left( r- r' \right) \ ,
\end{eqnarray}
where $\L$ is 
The bulk-to-bulk propagator is given by	
\begin{eqnarray}
\nonumber	G \left (r,t;r',t' \right) & = & \int \frac{d\omega}{2 \pi} e^{i \omega t } \left[\theta(r-r')~ f_1(r)f_2(r)+ \theta(r'-r)~f_1(r')f_2(r)\right] \\
\nonumber &=& \int \frac{d\omega}{2 \pi}  \left[\theta(r-r')~ e^{i\omega(t+r_*-t'-r_*')}+ \theta(r'-r)~e^{i\omega(t-r_*-t'+r_*')}\right]\\
\nonumber &=& \int \frac{d\omega}{2 \pi}  \left[\theta(r-r')~ e^{i\omega(v-v')}+ \theta(r'-r)~e^{i\omega(u-u')}\right]\\
	    &=& \theta(r-r') \delta (v-v') + \theta(r'-r) \delta (u-u') \ .
\end{eqnarray}
In the above, $r_*$ is the tortoise co-ordinate defined in (\ref{tortoiser}).
	
Now, near the boundary (\ref{solvecflucperp}) behaves as:
\begin{equation}
\delta A(t,r) \approx \left(C_1~ r^{\Delta _+}+ C_2 ~r^{\Delta_-}\right) e^{i \omega  t}  \ . 
\end{equation}
where $\Delta_+ =0$ and $\Delta_-=-1$. So the bulk-to -boundary propagator is given by
\begin{equation}
K(r,t;t') = \lim_{r'\rightarrow \infty}(\Delta_+ - \Delta_-)~ r^{\Delta_+} G(r,t;r',t') =  \delta\left(u-t + \frac{\pi  R^2 \omega }{4 r_{\rm E}}\right) \ . \label{btob1}
\end{equation}
Had we started with  Kernel function $e^{- i\omega(t-t')}$ then we would have had:
\begin{equation}
K(r,t;t')  = \delta\left(v-t - \frac{\pi  R^2 \omega }{4 r_{\rm E}}\right) \ . \label{btob2}
\end{equation}
For the quanta travelling along $v=0$ horizon we shall use (\ref{btob1}) whereas for the quanta moving along $u = 0$ we will use (\ref{btob2}). With these, the corresponding wave functions are obtained to be:
\begin{eqnarray}
\psi_1(p_1^U) &=& \frac{2 r_{\rm E}}{R^2} e^{\frac{\pi \omega }{2}} ~{\rm Exp} \left[\frac{2 r_{\rm E}}{R^2} t_1^{*} +  i ~ e^{-\frac{\pi}{2}}R^2~ p_1^U  ~e^{ \frac{2 r_{\rm E}}{R^2} t_1^{*} + \frac{\pi \omega}{2}}\right] \ , \\
\psi_2(p_2^V) &=&  \frac{2 r_{\rm E}}{R^2} e^{\frac{\pi \omega }{2}} ~{\rm Exp} \left[-\frac{2 r_{\rm E}}{R^2} t_2^{*} -  i~  e^{-\frac{\pi}{2}}R^2~ p_2^V  ~e^{- \frac{2 r_{\rm E}}{R^2} t_2^{*} + \frac{\pi \omega}{2}}\right] \ , \\
\psi_3(p_3^U) &=& \frac{2 r_{\rm E}}{R^2} e^{\frac{\pi \omega }{2}} ~{\rm Exp} \left[\frac{2 r_{\rm E}}{R^2} t_3 +  i ~ e^{-\frac{\pi}{2}}R^2~ p_1^U  ~e^{ \frac{2 r_{\rm E}}{R^2} t_3 + \frac{\pi \omega}{2}}\right] \ , \\ 
\psi_4(p_4^V) &=&  \frac{2 r_{\rm E}}{R^2} e^{\frac{\pi \omega }{2}} ~{\rm Exp} \left[-\frac{2 r_{\rm E}}{R^2} t_4 -  i ~ e^{-\frac{\pi}{2}}R^2~ p_2^V  ~e^{- \frac{2 r_{\rm E}}{R^2} t_4 + \frac{\pi \omega}{2}}\right] \ .
\end{eqnarray}
Now, it is straightforward to use these wave functions to construct of four-point OTOC, by computing the overlap integral.

\subsection{The Overlap Integral}

To evaluate the overlap integral, let us make the following change of variables: 
\begin{eqnarray}
&& p_1^U = \frac{p}{i e^{-\frac{\pi}{2}} R^2 \left({\rm exp} \left({\frac{2 r_{\rm E}}{R^2} t_3^{*}}\right) - {\rm exp} \left({\frac{2 r_{\rm E}}{R^2} t_1^{*}} \right)\right)}  \ , \\
&& p_2^V = \frac{q}{i e^{-\frac{\pi}{2}} R^2 \left( {\rm exp} \left({-\frac{2 r_{\rm E}}{R^2} t_2^{*}} \right) - {\rm exp} \left({-\frac{2 r_{\rm E}}{R^2} t_4^{*}} \right) \right)} 
\end{eqnarray}
and also, analytically continue to $ t_1 = i \epsilon_1$ , $t_2 = t + i \epsilon_2$, $ t_3 = i \epsilon_3$, $t_4 = t + i \epsilon_4$. This yields:
\begin{eqnarray}
&& \nonumber \langle A( t+ i \epsilon_4) A(i \epsilon_3) A(t+ i \epsilon_2)  A(i \epsilon_1)  \rangle = \mathcal C \int_0^{\infty} dp dq \left( p  q\right)  e^{-p-q} I\left( p, q\right) \ ,  \nonumber\\
&& I\left( p, q\right)  =  {\rm exp} \left[ {i \left( p q \right) e^{\frac{2 r_{\rm E}}{R^2} t} ~ \frac{N e^{\frac{3 \pi}{2}}}{2 R^6 r_{\rm E}^2 \epsilon_{13}^{*} \epsilon_{24} T_{{\rm D}5} }} \right] ,\\ 
&& \langle A(i \epsilon_3)   A(i \epsilon_1)  \rangle \langle A(t+ i \epsilon_4)  A(t+ i \epsilon_2)    \rangle =  \mathcal C \int_0^{\infty}  dp dq \left( p  q \right)  e^{-p-q}  \ .
\end{eqnarray}
In the above, we have also defined:
\begin{eqnarray}
&& \epsilon_{ij} = i \left( e^{ \frac{2 i r_{\rm E}}{R^2} \epsilon_i} - e^{ \frac{2 i r_{\rm E}}{R^2} \epsilon_i}\right) \  , \\
&& \mathcal{C} = \frac{256 r_{\rm E}^4~ e^{2 \pi \omega}~ {\rm exp} \left[ {\frac{2 r_{\rm E}}{R^2} \left(- i \epsilon_1+ i \epsilon_2 -i \epsilon_3 +i \epsilon_4\right)} \right]}{\pi^2  e^{-2\pi} R^{16} \left(\epsilon_{13}^* \right)^2 \left(\epsilon_{24} \right)^2 } \ .
\end{eqnarray}
 Finally, the normalised OTOC is given by
\begin{equation}
\frac{\langle A(t+ i \epsilon_4) A(i \epsilon_3) A(t+ i \epsilon_2)  A(i \epsilon_1)  \rangle}{\langle A(i \epsilon_3)   A(i \epsilon_1)  \rangle \langle A(t+ i \epsilon_4)  A(t+ i \epsilon_2)    \rangle} = 1+ \frac{i  N e^{\frac{3 \pi}{2}}}{2 R^6 r_{\rm E}^2 \epsilon_{13}^{*} \epsilon_{24} ~ T_{{\rm D}5}  } e^{\frac{2 r_{\rm E}}{R^2} t} \ . \label{otocnorm}
 \end{equation}

Recall that, the temperature associated with the osm event horizon is given by
\begin{equation}
T = \frac{\pi r_{\rm E}}{R^2} = \frac{1}{\beta} \ . \label{d5T}
\end{equation}
Now, we make the following choices for imaginary time parameters\cite{Shenker:2014cwa}: $\epsilon_1= 0$, $\epsilon_2= -\frac{\beta}{4}$ , $\epsilon_3= \frac{\beta}{2}$, $\epsilon_4= \frac{\beta}{4}$. With these, and restoring the temperature dependence, (\ref{otocnorm}) becomes:
\begin{equation}
\frac{\left<A(t+ i \epsilon_4) A(i \epsilon_3) A(t+ i \epsilon_2)  A(i \epsilon_1)  \right>}{\left< A(i \epsilon_3)   A(i \epsilon_1)  \right>\left<A(t+ i \epsilon_4)  A(t+ i \epsilon_2)    \right>} = 1 -  \frac{ ~ N e^{\frac{3 \pi}{2}}}{8 R^6 r_{\rm E}^2  ~ T_{{\rm D}5} } e^{\frac{2 \pi}{\beta} t} \ .
\end{equation}
Therefore, we finally recover a maximal chaos on the D-brane horizon: $\lambda_{\rm L} = 2 \pi T$, while the temperature is given by (\ref{d5T}).

A few comments are in order. First, note that the scrambling time behaves as: 
\begin{eqnarray}
t_{\rm sc} \sim \frac{1}{2\pi T} \log \left( \sqrt{\lambda} N_c\right) \ , 
\end{eqnarray}
which is obtained by substituting the D$5$-brane tension in terms of string length and string coupling. This result is consistent with the anticipated answer in (\ref{tscgen}). Moreover, the above calculation can be explicitly carried out when there is an event horizon in the background AdS$_5$ geometry. Suppose the corresponding background temperature is given by $T$, and the electric field on the brane is given by $E$, then the resulting Lyapunov exponent is given by
\begin{eqnarray}
\lambda_{\rm L} = 2 \pi \left( T^4 + E^2 \right)^{1/4} \ .
\end{eqnarray}
Thus the brane degrees of freedom have a larger Lyapunov exponent compared to the gravity degrees of freedom. This is due to the lack of strict thermal equilibrium in this configuration, and the probe sector being an open system.

\section{General Soft Sector}

In this section, we will discuss the computation of a $4$-point OTOC and the corresponding exponential growth of the same, for an one-dimensional system with soft degrees of freedom. This is, presumably, determined by a general functional of the Schwarzian derivative. Here we will not address the question of the UV-theory from which such an effective description may emerge. Therefore, our analyses and conclusions remain valid as a strict IR-physics. 

Schematically, a general action is given by
\begin{eqnarray}
&& S_{\rm IR} = - \frac{1}{g^2} \int dt \F \left[ \{\varphi(t), t \} \right] \ , \quad {\rm with} \nonumber\\
&& \{\varphi(t), t \} = \frac{ - 3 \ddot{\varphi} ^2 + 2 \dot{\varphi}  \dddot{\varphi}}{2 \dot{\varphi}^2} \ . \label{IRgen}
\end{eqnarray}
Here, $g$ is the corresponding coupling and $t$ is the Euclidean time. The Lorentzian time is related to the Euclidean time {\it via}: $t =  i t_{\rm L}$. The global symmetry of the action (\ref{IRgen}) is associated with the following SL$(2,R)$ transformation:
\begin{eqnarray}
\varphi(t) = \frac{a \varphi(t) + b}{c \varphi(t) + d } \ .
\end{eqnarray}
The general equation of motion obtained from the above equation is given by
%
%
%
\begin{eqnarray}
&& \partial_t \left[ - \partial_t^2 \left( \frac{\F '}{\dot{\varphi}} \right) + \F' \left( \frac{\dddot{\varphi}}{\dot{\varphi}^2} - 3 \frac{\ddot{\varphi}^2}{\dot{\varphi}^3} \right) - 3 \partial_t \left( \F' \frac{\ddot{\varphi}}{\dot{\varphi}^2} \right) \right] = 0 \ , \label{eqngen} \\
&& \F ' \equiv \frac{\partial \F}{\partial y} \ , \quad y \equiv  \{\varphi(t), t \} \ , 
\end{eqnarray}
The above is a sixth-order differential equation. It is easy to check that for $\F(y) = y$, we get back the standard quartic equation of motion $\partial_t \{\varphi(t), t \} = 0$. It is also straightforward to check that any solution of the purely Schwarzian derivative equation of motion is also a solution of (\ref{eqngen}), irrespective of the functional form of $\F$. There can be special and new solutions of (\ref{eqngen}), which fall outside the solution space spanned by the solutions of the $\partial_t \{\varphi(t), t \} = 0$ equation; however, we will not explore those cases.

In order to analyze the equations of motion in more explicit terms, let us use the co-ordinate transformation\cite{Maldacena:2016upp}: $\varphi = \tan \left( \psi/2 \right)$ with $\F = \{\varphi(t), t \}^n$, $n \in {\mathbb Z}_+$. The action in (\ref{IRgen}) takes the following form:
\begin{eqnarray}
S_{\rm IR} = - \frac{1}{g^2} \int dt \left[ \left( \{\psi(t), t \} \right) + \frac{1}{2} \left(\partial_t \psi\right)^2 \right]^n \ , \label{irgenpsi}
\end{eqnarray}
Using 
\begin{eqnarray}
\psi(t) = t + \epsilon(t) \ , 
\end{eqnarray}
the action, at the leading order, is given by
\begin{eqnarray}
S_{\rm IR} =   - \frac{ n}{ 2^{n} g^2} \int dt \left[   \left(2 (n-1) \epsilon ^{(3)}(t)^2+(2 n-1) \epsilon '(t)^2+2 (2 n-3) \epsilon ^{(3)}(t) \epsilon '(t)-3 \epsilon ''(t)^2\right) \right] \ . \nonumber \\ 
\end{eqnarray}
The corresponding equation of motion is given by
\begin{eqnarray}
4 (n-1) \epsilon ^{(6)}(t) + 4 (2 n-3) \epsilon ^{(4)}(t) + 2 (2 n-1) \epsilon ^{(2)}(t) + 6 \epsilon ^{(4)}(t) = 0 \ .
\end{eqnarray}
The zero modes of this equation, on the solution space $\epsilon(t) = e^{i k t}$, are given by
\begin{eqnarray}
k = 0 \ , \, \pm 1 \ , \, \pm \frac{\sqrt{2 n-1}}{\sqrt{2 n-2}} \ .
\end{eqnarray}
In above, we have assumed that $\frac{\sqrt{2 n-1}}{\sqrt{2 n-2}}$ is an integer, which restricts the allowed space for $n$. Clearly, the modes $k =  0, \pm 1$ is present irrespective of the value of $n$. Typically, the presence of the additional zero modes imply there is a bigger symmetry, compared to an SL$(2,R)$; however, it is not clear to us what physical symmetry this corresponds to and we will not elaborate on this. The corresponding propagator is given by
\begin{eqnarray}
G (t) =  \sum_k e^{i 2\pi k t / \beta}G_k \ ,  \quad G_k = \frac{1}{k^2 \left( k^2 -1 \right) \left( k^2 - k_n^2 \right) } \ , \quad k_n = \frac{\sqrt{2 n-1}}{\sqrt{2 n-2}} \ .
\end{eqnarray}
In the above sum, we need to avoid the zero modes. The Matsubara summation can be carried out, using the following integral:
\begin{eqnarray}
G(t) = \int_{\cal C} d\omega e^{(2\pi t \omega)/\beta} \frac{2\pi}{e^{2\pi\omega} -1 } \frac{1}{\omega^2 \left( \omega^2 + 1\right ) \left( \omega^2 + k_n^2 \right) } \ ,
\end{eqnarray}
where the contour, ${\cal C}$, is chosen such that the poles at the $\omega = - i \infty, \ldots + i \infty$ are included, except $\omega = 0, \pm i, \pm k_n$. Clearly, the integrand goes to zero as $|\omega| \to \infty$ for $t < 0$. To ensure a similar convergence, we can simply consider an integrand in which the exponential has a reversed time direction. Thus, all contributions to the above integral, from infinity, will be zero.

\begin{figure}[ht!]
\begin{center}
{\includegraphics[width=0.8\textwidth]{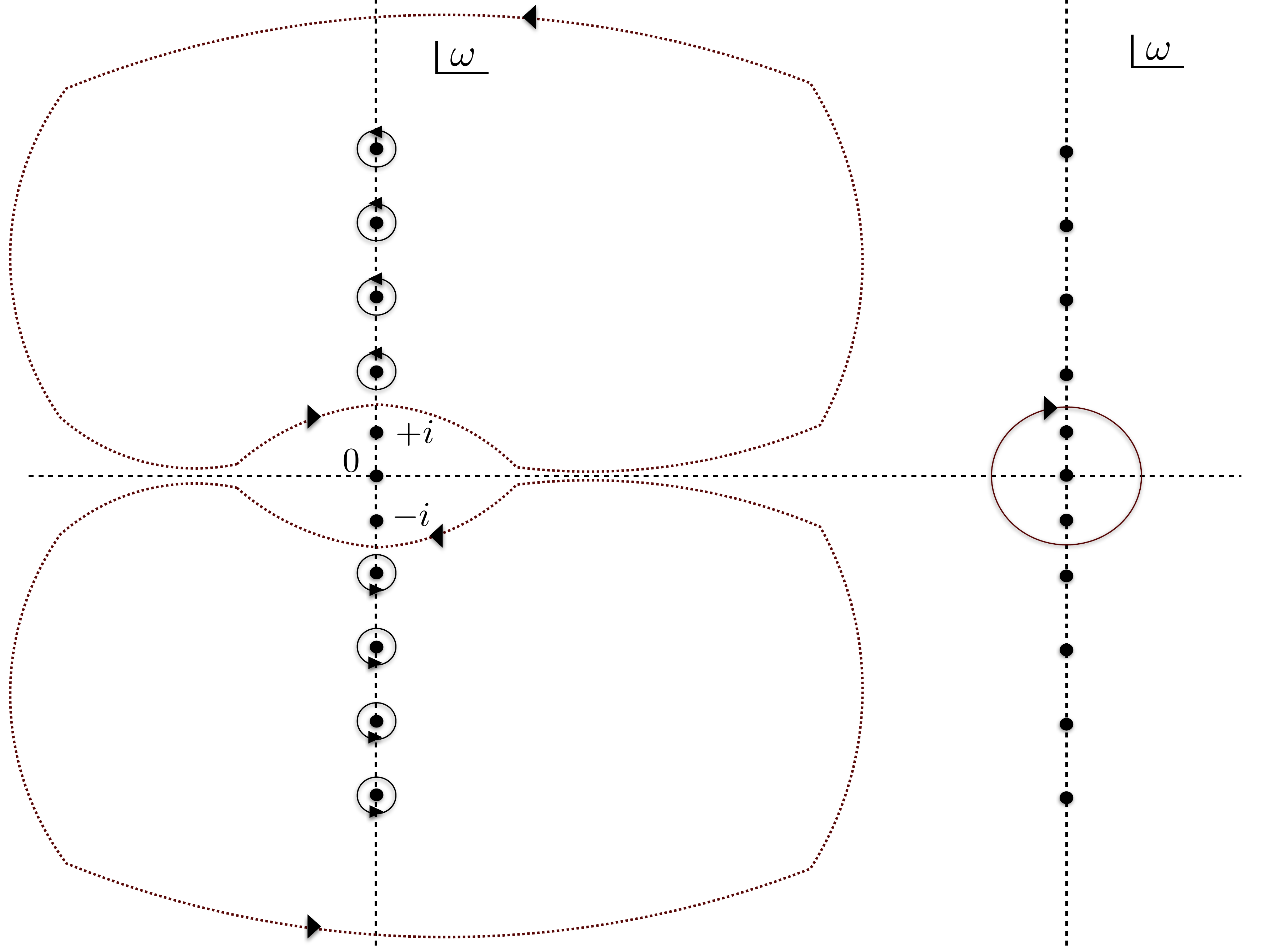}}
\caption{\small This is a pictorial representation of performing the Matsubara summation {\it via} a contour integration. On the left, one gathers the residues at each pole, except the zero modes. Using the bounded-ness of the integrand as $|\omega| \to \infty$, we can use an alternative contour, denoted by the dashed curve, which collects all the residues. When a similar contour is drawn in the lower half plane, the combined contour can equivalently be thought of as a contour around the zero-modes. } \label{zeromodes}
\end{center}
\end{figure}

In turn, the integral now boils down to a sum of contour integrals around $\omega = 0, \pm i, \pm k_n$, which is pictorially demonstrated in figure \ref{zeromodes}. 
The resulting propagator is given by
\begin{eqnarray}
G(t) = && \frac{1}{6 k_n^4 \left( k_n^2 - 1\right)^2} \left[ ( k_n^2 - 1) \left( 6 k_n^4 (\pi - |t|) \sin |t| + \right. \right. \nonumber\\
&& \left. \left.  \left( k_n^2 - 1\right) \left(k_n ^2 \left( 3 |t|^2 - 6 \pi  |t| + 2 \pi^2 - 6 \right) - 6 \right) \right. \right. \nonumber\\
&& \left. \left. - 6 \pi k_n \csc (\pi k_n) \cos ( k_n (\pi - |t| ) ) \right) + 3 \left( 9 - 5 k_n^2\right) k_n^4 \cos |t| \right]   \ , \label{propn} 
\end{eqnarray}
To make connection with already known result in \cite{Maldacena:2016upp}, when $n=1$ in (\ref{IRgen}), the propagator is given by
\begin{eqnarray}
G(t) = \frac{1}{6} \left(3 |t|^2-6 \pi  |t| +6 (\pi - |t| ) \sin |t| - 15 \cos |t| + 2 \pi ^2 - 6 \right) \ . \label{prop1}
\end{eqnarray}
In both the above expressions, we have ignored an overall constant. Note that, starting from the expression in (\ref{propn}), one cannot obtain (\ref{prop1}) by simply substituting $n=1$, since this is a singular substitution. Hence, the $n=1$ case needs to be worked out separately.

Now, similar to \cite{Maldacena:2016upp}, we want to evaluate the four point function $\langle V (\varphi_1) V(\varphi_2) W(\varphi_3) W(\varphi_4)\rangle$. Using a large $N$ factorization, the leading large $N$ answer is given by 
\begin{eqnarray}
\langle V (\varphi_1) V(\varphi_2) W(\varphi_3) W(\varphi_4)\rangle \sim \left( \frac{1}{\varphi_{12}} \right)^{2\Delta_1} \left( \frac{1}{\varphi_{34}} \right)^{2\Delta_2} \ , \quad \varphi_{ij} \equiv \varphi_i - \varphi_j \ , \label{fourpoint}
\end{eqnarray}
where $\Delta_1$ and $\Delta_2$ are the conformal dimensions of the field $V$ and $W$, respectively. Under the following transformaiton
\begin{eqnarray}
\varphi = \tan \left( \frac{\psi}{2}\right) \ , \quad \psi(t) = t + \epsilon(t) \ ,
\end{eqnarray}
we get
\begin{eqnarray}
\langle V (\varphi_1) V(\varphi_2) W(\varphi_3) W(\varphi_4)\rangle \sim  &&\left[ \frac{ \left(1 + \epsilon' (t_1) \right)^{\Delta_1} \left(1 + \epsilon' (t_2) \right)^{\Delta_1}}{ \left( \sin \left( \frac{t_{12} + \epsilon(t_1) - \epsilon(t_2) }{2} \right)\right)^{2\Delta_1} }  \right] \nonumber\\
&& \left[\frac{ \left(1 + \epsilon' (t_3) \right)^{\Delta_2} \left(1 + \epsilon' (t_4) \right)^{\Delta_2}}{ \left( \sin \left( \frac{t_{34} + \epsilon(t_3) - \epsilon(t_4) }{2} \right)\right)^{2\Delta_2} }  \right]\ , 
\end{eqnarray}
which subsequently yields:
\begin{eqnarray}
&& \C_{(4)} \equiv \frac{\langle V (t_1) V(t_2) W(t_3) W(t_4)\rangle}{\langle V (t_1) V(t_2) \rangle \langle W(t_3) W(t_4)\rangle}  \nonumber\\
&& \sim  1 + \Delta_1 \Delta_2 \left( \frac{\epsilon (t_1) - \epsilon (t_2)}{ \tan \left( \frac{t_{12}}{2}\right) } - \epsilon '( t_1) - \epsilon '(t_2)\right) \left( \frac{\epsilon (t_3) - \epsilon (t_4)} {\tan \left( \frac{t_{34}}{2}\right)} - \epsilon'(t_3) - \epsilon '(t_4)\right) \ . \label{4ptform} \nonumber\\
\end{eqnarray}
The formula above is already explicitly presented in \cite{Maldacena:2016upp}. When the low energy description is a non-trivial functional of Schwarzian derivative, we can now use the propagator in (\ref{propn}), which, for $n=1$, reduces to (\ref{prop1}). We will use the explicit form of the propagator with the following identifications:
\begin{eqnarray}
&& \epsilon(t_1) \epsilon(t_2) \equiv G(t_1 - t_2) \ , \quad t_1 > t_2 \ ,  \label{propad} \\
&& \epsilon(t_1) \epsilon(t_2) \equiv G( - t_1 + t_2) \ , \quad t_1 < t_2 \ . \label{propret}
\end{eqnarray}

Thus, we need to evaluate the $\epsilon$-dependent terms in (\ref{4ptform}), with a particular time-ordering. The time-ordering is taken care by the definitions in (\ref{propad})-(\ref{propret}), and the explicit function $G(t)$ is given in (\ref{propn}). Let us first evaluate the correlator, which yields:
\begin{eqnarray}
\C_{(4)} - 1 \sim  && \cot \left( \frac{ t_1 - t_2 }{2} \right) \left( - G^{(0,1)}(t_1, t_3) - G^{(0,1)}( t_1 , t_4) + G^{(0,1)}( t_2 , t_3) + G^{(0,1)}(t_2 , t_4)\right) \nonumber\\
&& + \cot \left( \frac{t_3 - t_4}{2} \right) \left( - G^{(1,0)}( t_1 , t_3) + G^{(1,0)}(t_1 , t_4) - G^{(1,0)}(t_2, t_3) + G^{(1,0)}( t_2 , t_4) \right) \nonumber\\
&& +  G^{(1,1)}(t_1 , t_3) + G^{(1,1)}(t_1 , t_4) + G^{(1,1)}(t_2 , t_3) + G^{(1,1)}(t_2 , t_4) \nonumber\\
&& + \cot \left(\frac{t_1 - t_2}{2}\right) \cot \left(\frac{t_3 - t_4}{2}\right) (G(t_1 , t_3) - G(t_1 , t_4) - G(t_2, t_3) + G(t_2 , t_4)) \label{4ptgen}  \nonumber\\
\end{eqnarray}
Depending on the theory, one can now directly substitute the propagator in (\ref{4ptgen}). The information of the time-ordering is operationally implemented in the absolute value of the time-separation in each propagator, as is explicitly written in (\ref{propn}) and (\ref{prop1}). To evaluate the time-ordered correlator, we impose the following ordering: $t_1 > t_2 > t_3 > t_4$. Similarly, the four-point OTO correlator is evaluated with the time ordering: $t_1 > t_3 > t_2 > t_4$.

In the former case, with $t_1 > t_2 > t_3 > t_4$, one does not observe any exponential growth or the correlator, for any generic value of $n$, and we do not elaborate on this case any further. Now, with $t_1 > t_3 > t_2 > t_4$, and substituting $t_1 = i a$, $t_2 = 0 $, $t_3 = i (t_{\rm L} + b)$ and $t_4 = i t_{\rm L}$, we get:
\begin{eqnarray}
\C_{(4)} - 1 \sim && - i \coth \left(\frac{a}{2}\right) \left( - G^{(0,1)}(i a , i ( b + t_{\rm L} )) - G^{(0,1)}(i a , i t_{\rm L} ) +  \right. \nonumber\\
&& \left. G^{(0,1)}(i (b + t_{\rm L} ), 0) + G^{(0,1)}(0,i t_{\rm L} )\right) + \nonumber\\
&& i \coth \left( \frac{b}{2}\right) \left( G^{(1,0)}(i a , i (b + t_{\rm L})) - G^{(1,0)}( i a , i t_{\rm L} ) + G^{(1,0)}(  i ( b + t_{\rm L} ), 0) - G^{(1,0)}(0 , i t_{\rm L})\right) \nonumber\\
&& + \coth \left(\frac{a}{2}\right) \coth \left(\frac{b}{2}\right) (- G(i a, i (b + t_{\rm L} )) + G( i a , i t_{\rm L} ) + G(i ( b + t_{\rm L} ), 0) - G(0 , i t_{\rm L} )) \nonumber\\
&&  + G^{(1,1)}( i a, i (b + t_{\rm L} )) + G^{(1,1)}(i a , i t_{\rm L} ) + G^{(1,1)}( i (b + t_{\rm L} ), 0 ) + G^{(1,1)}(0, i t_{\rm L}) \ .
\end{eqnarray}
Now, using the explicit propagator in (\ref{propn}), it is easy to check that the leading behaviour in the real time coordinate $t_{\rm L}$ comes from the leading trigonometric function after performing analytic continuation. The relevant pieces are: $\sin |t|$, $\cos |t|$ and $\cos \left( k_n \pi - k_n |t|\right) $. Each of these pieces yields an $e^{t_{\rm L}}$ or $e^{k_n t_{\rm L}}$ factor after the analytic continuation above. In the limit $t_{\rm L} \gg \{a, b\}$, this exponential piece will dominate the corresponding expression. This, schematically yields:\footnote{Note that, it is possible, for certain values of $n$, the coefficient of the exponential growth vanishes and therefore the correlator has no growth with an $e^{k_n t_{\rm L}}$ mode. However, this is not a generic situation and may happen for fine-tuned values of $n$. We will not explore this possibility here.}
\begin{eqnarray}
\C_{(4)} - 1  \sim  e^{ \lambda_{\rm L} t_{\rm L}} + {\rm sub-leading} \ ,
\end{eqnarray}
with 
\begin{eqnarray}
\lambda_{\rm L} = \frac{2\pi}{\beta} {\rm max} \left(1, k_n \right) \ ,
\end{eqnarray}
where the explicit thermal factor of $(2\pi)/ \beta$ has been restored. The ${\rm max}$ function selects the greatest real number in the list. In the regime $1 \ge k_n$, we obtain a canonical chaos bound saturation: $\lambda_{\rm L} = 2\pi T$. On the other hand, when $k_n > 1$, we obtain:
\begin{eqnarray}
\lambda_{\rm L}  = \frac{2\pi}{\beta} k_n = \frac{2\pi}{\beta} \frac{\sqrt{2n -1}}{\sqrt{2 n - 2 }} \ . \label{lyan}
\end{eqnarray}
Before moving on, let us make a simple observation. In \cite{Perlmutter:2016pkf}, a corresponding chaos-bound was explored for a two-dimensional conformal field theory with higher spin currents, with a bound on the spin $ s \le N $. Here $N$ is a finite number and greater than two. In this case, the corresponding Lyapunov exponent is given by $\lambda_{\rm L} = (2\pi T) (N-1)$. A simple algebraic map $n = (1 - 2 N^2) / (2 - 2 N^2)$ directly relates this to the formula in (\ref{lyan}). In particular, for $N=3$, one obtains $n = 1 + \frac{1}{16}$.

The action functional ${\cal F}$ has imprints on the scrambling time as well, denoted by $t_{\rm sc}$. In the case $k_n > 1$, this is determined by
\begin{eqnarray}
\cO \left( {\cal K}(n) e^{ \frac{2\pi}{\beta} k_n t_{\rm sc}} \right) = \cO (1) \quad \implies \quad t_{\rm sc} \sim \frac{1}{2\pi T k_n} \log \left( {\cal K}(n) \right) \ .
\end{eqnarray}
Here ${\cal K}(n)$ is a function of the index $n$ as well as the central charge of the system. The corresponding scrambling time is further suppressed. Evidently, in the case $k_n < 1$, the scrambling time is the standard $t_{\rm sc} \sim \beta / (2 \pi) \log ({\rm central \, charge})$. As remarked in \cite{Perlmutter:2016pkf}, the violation of the chaos bound or the {\it too-fast} scrambling (for $k_n >1$) indicates potential problems with such theories. The $k_n <1 $ region is therefore, {\it a priori}, problem-free. This corresponds to $n \in (0, \frac{1}{2}]$, for positive values of $n$. Clearly, $n=1$ is the standard Schwarzian theory, which does not fall under the above analysis. Note also that, when $k_n$ is purely imaginary, the additional poles in figure \ref{zeromodes} sit on the real $\omega$ axis (the horizontal axis). Correspondingly, the contour in the upper half plane in figure \ref{zeromodes} can be appropriately deformed to avoid those. In turn, once again the resulting sum will be obtained by evaluating the residues around the zero modes. This has no bearing on the exponential growth of the four point OTOC, since, upon subsequent analytic continuation, these zero modes will contribute to oscillatory behaviour. This corresponds to the parametric space $n \in [\frac{1}{2}, 1]$. Incidentally, the entire $n < 0$ regime is also allowed by this consideration. Finally, note that, in the $n \to \infty$ limit, $k_n \to 1$, and thus the point at infinity also seems viable. However, this is a more subtle point since one needs to investigate how to close the contour at infinity when such a pole is present. Summarizing, the allowed range of the power in (\ref{IRgen}) is given by
\begin{eqnarray}
n \in \left(-\infty,  0 \right) \cup \left( 0 , 1 \right] \ ,
\end{eqnarray}
in which $n$ takes values such that the zero modes occur at integers values.

Before concluding this section let us briefly look at the thermodynamic properties of the effective action (\ref{IRgen}). We have chosen the overall sign in front of the integral in (\ref{IRgen}) in congruence with \cite{Maldacena:2016upp}, such that the $n=1$ case readily reproduces the standard Schwarzian action. The general equation of motion can be derived from (\ref{irgenpsi}), which has a simple solution\cite{Maldacena:2016upp}: $\psi = (2\pi / \beta) t$. The factor of $(2\pi/\beta)$ is chosen such that $\psi = \psi + 2\pi$, upon the identification $t = t + \beta$. On this solution, the on-shell action can be evaluated which yields the logarithm of the corresponding partition function:
\begin{eqnarray}
\log Z = \frac{2^n \pi^{2 n}}{g^2} \beta^{1-2n} \ ,
\end{eqnarray}
which yields the following entropy:
\begin{eqnarray}
s = \log Z - \beta \frac{\partial \log Z}{\partial \beta} = \frac{2^{n+1} n \pi ^{2 n} \beta ^{1-2 n}}{g^2} \ .
\end{eqnarray}
The corresponding specific heat scales as $c_V \propto T^{2n-1}$, where the proportionality constant is positive. The sign of the proportionality constant is directly related to the sign of the action in (\ref{IRgen}). Hence $c_V > 0$ is guaranteed by design. Note, however, that for $n=1/2$, the specific heat becomes temperature independent. The entropy, therefore, can be completely attributed to be an extremal entropy. Moreover, for $n<1/2$, the corresponding temperature scaling is rather unusual.

\section{Conclusions}

In this article we have explored a few aspects related to maximal chaos. First, we obtained an effective Schwarzian action from an open string embedded in an AdS-background, including the interaction term of this Schwarzian modes with a general fluctuation degree of freedom. Secondly, we explicitly demonstrated a few examples from a D-brane worldvolume perspective that, even without a strictly gravitational theory, maximal chaos can be obtained from a non-linear theory, such as a DBI-action. In fact, although we have not included the explicit example, one can consider more general backgrounds in which a similar analytical calculations can be performed. For example, take the so-called Lifshitz symmetric geometries, specially when the dynamical exponent is equal to two. Following \cite{Kundu:2015qda}, the gauge field fluctuations around a space-filling defect brane-like degree of freedom can be analyzed, similar to the D$5$-brane fluctuations that we have described in details. For this case, also, one observes a maximal chaos set by the constant electric field on the brane worldvolume. Finally, we have also demonstrated that for a general class of systems, described by a functional of the Schwarzian action, maximal chaos will occur.

We have left many stones unturned. For example, it is likely that a purely worldsheet analysis exists in describing the Schwarzian action and its' coupling to other heavier modes. This will be a very interesting aspect to chalk out. More speculatively, a similar description may also exist for a general D-brane worldvolume, perhaps not in terms of a Schwarzian effective dynamics. In such cases, the corresponding Rindler dynamics dictates the maximal chaos and it is an interesting possibility if there exists an effective description of a soft sector around a Rindler geometry. In the context of a general semi-classical system, which may or may not have a holographic dual, it is also an interesting question to explore whether maximal chaos selects out a special class of systems, see {\it e.g.}~\cite{Roberts:2014ifa, Fitzpatrick:2016thx} for some related studies in CFTs.

There are certainly relatively straightforward follow-ups of the current work. For example, a comprehensive analysis of the fluctuation modes including the longitudinal fluctuations will complete the discussion. The apparent violation of the chaos bound for the D-brane fluctuations is intuitively due to the lack of a precise thermal equilibrium in the configuration, however, it will be useful to figure out precisely what step of \cite{Maldacena:2015waa} is violated by such a D-brane configuration.

Generically, it seems non-linearity is essential for the exponential growth of the corresponding OTOC, however, it remains unclear whether there is a special class of such non-linear theories. A four-point OTOC generally grows in time, but with a power law. It would be rather interesting to sharpen this difference, even on a class of examples. Moreover, one may investigate higher point OTOCs, although these are generally hard to calculate, except in relatively simple cases such as the IR-limit of the Sachdev-Ye-Kitaev-type model itself\cite{Bhattacharya:2018nrw}.\footnote{See {\it e.g.}~\cite{Mandal:2017thl, Gaikwad:2018dfc} towards a gravity construction of the SYK-model. See also \cite{Sarosi:2017ykf} for an excellent review on the SYK-physics and its connection with AdS$_2$-physics.} Another intriguing aspect is the nature of growth of an OTOC in a generic state, which is not thermal. Certainly, from a classical perspective, one does not require a thermal state to define a Lyapunov exponent, and there may be an analogue in the semi-classical scenario. We hope to address some of these issues in near future.

\section{Acknowledgements}

We thank Jan de Boer, Gautam Mandal, Shiraz Minwalla, Juan F.~Pedraza, Shibaji Roy, Harvendra Singh for various conversations. A special thanks goes to Gautam Mandal and Juan F.~Pedraza for patiently clarifying many issues of the OTOC calculation.


\begin{thebibliography}{99}


\bibitem{Gubser:1998bc} 
  S.~S.~Gubser, I.~R.~Klebanov and A.~M.~Polyakov,
  ``Gauge theory correlators from noncritical string theory,''
  Phys.\ Lett.\ B {\bf 428}, 105 (1998)
  doi:10.1016/S0370-2693(98)00377-3
  [hep-th/9802109].


\bibitem{Witten:1998qj} 
  E.~Witten,
  ``Anti-de Sitter space and holography,''
  Adv.\ Theor.\ Math.\ Phys.\  {\bf 2}, 253 (1998)
  doi:10.4310/ATMP.1998.v2.n2.a2
  [hep-th/9802150].
  
  
\bibitem{Herzog:2002pc} 
  C.~P.~Herzog and D.~T.~Son,
  ``Schwinger-Keldysh propagators from AdS/CFT correspondence,''
  JHEP {\bf 0303}, 046 (2003)
  doi:10.1088/1126-6708/2003/03/046
  [hep-th/0212072].
  
\bibitem{SonnerLecture}
http://www.sonnerphysics.net/uploads/6/1/9/9/61993493/solvayfull.pdf


\bibitem{Shenker:2013pqa} 
  S.~H.~Shenker and D.~Stanford,
  ``Black holes and the butterfly effect,''
  JHEP {\bf 1403}, 067 (2014)
  doi:10.1007/JHEP03(2014)067
  [arXiv:1306.0622 [hep-th]].
  

\bibitem{Shenker:2013yza} 
  S.~H.~Shenker and D.~Stanford,
  ``Multiple Shocks,''
  JHEP {\bf 1412}, 046 (2014)
  doi:10.1007/JHEP12(2014)046
  [arXiv:1312.3296 [hep-th]].


\bibitem{Shenker:2014cwa} 
  S.~H.~Shenker and D.~Stanford,
  ``Stringy effects in scrambling,''
  JHEP {\bf 1505}, 132 (2015)
  doi:10.1007/JHEP05(2015)132
  [arXiv:1412.6087 [hep-th]].
  
  
\bibitem{Maldacena:2015waa} 
  J.~Maldacena, S.~H.~Shenker and D.~Stanford,
  ``A bound on chaos,''
  JHEP {\bf 1608}, 106 (2016)
  doi:10.1007/JHEP08(2016)106
  [arXiv:1503.01409 [hep-th]].
  
  
\bibitem{Banerjee:2018twd} 
  A.~Banerjee, A.~Kundu and R.~R.~Poojary,
  ``Strings, Branes, Schwarzian Action and Maximal Chaos,''
  arXiv:1809.02090 [hep-th].
  

\bibitem{deBoer:2017xdk} 
  J.~de Boer, E.~Llabr\'{e}s, J.~F.~Pedraza and D.~Vegh,
  ``Chaotic strings in AdS/CFT,''
  Phys.\ Rev.\ Lett.\  {\bf 120}, no. 20, 201604 (2018)
  doi:10.1103/PhysRevLett.120.201604
  [arXiv:1709.01052 [hep-th]].
  

\bibitem{Murata:2017rbp} 
  K.~Murata,
  ``Fast scrambling in holographic Einstein-Podolsky-Rosen pair,''
  JHEP {\bf 1711}, 049 (2017)
  doi:10.1007/JHEP11(2017)049
  [arXiv:1708.09493 [hep-th]].
  

\bibitem{Cai:2017nwk} 
  R.~G.~Cai, S.~M.~Ruan, R.~Q.~Yang and Y.~L.~Zhang,
  ``The String Worldsheet as the Holographic Dual of SYK State,''
  arXiv:1709.06297 [hep-th].
  
  
\bibitem{Xiao:2008nr} 
  B.~W.~Xiao,
  ``On the exact solution of the accelerating string in AdS(5) space,''
  Phys.\ Lett.\ B {\bf 665}, 173 (2008)
  doi:10.1016/j.physletb.2008.06.017
  [arXiv:0804.1343 [hep-th]].
  
  
\bibitem{Jensen:2013ora} 
  K.~Jensen and A.~Karch,
  ``Holographic Dual of an Einstein-Podolsky-Rosen Pair has a Wormhole,''
  Phys.\ Rev.\ Lett.\  {\bf 111}, no. 21, 211602 (2013)
  doi:10.1103/PhysRevLett.111.211602
  [arXiv:1307.1132 [hep-th]].
  
  
\bibitem{Polyakov:1987zb} 
  A.~M.~Polyakov,
  ``Quantum Gravity in Two-Dimensions,''
  Mod.\ Phys.\ Lett.\ A {\bf 2}, 893 (1987).
  doi:10.1142/S0217732387001130
  

\bibitem{Maldacena:2016upp} 
  J.~Maldacena, D.~Stanford and Z.~Yang,
  ``Conformal symmetry and its breaking in two dimensional Nearly Anti-de-Sitter space,''
  PTEP {\bf 2016}, no. 12, 12C104 (2016)
  doi:10.1093/ptep/ptw124
  [arXiv:1606.01857 [hep-th]].
  
  
\bibitem{Jensen:2016pah} 
  K.~Jensen,
  ``Chaos in AdS$_2$ Holography,''
  Phys.\ Rev.\ Lett.\  {\bf 117}, no. 11, 111601 (2016)
  doi:10.1103/PhysRevLett.117.111601
  [arXiv:1605.06098 [hep-th]].
  
  
\bibitem{Engelsoy:2016xyb} 
  J.~Engels\"{o}y, T.~G.~Mertens and H.~Verlinde,
  ``An investigation of AdS$_{2}$ backreaction and holography,''
  JHEP {\bf 1607}, 139 (2016)
  doi:10.1007/JHEP07(2016)139
  [arXiv:1606.03438 [hep-th]].
  
  
\bibitem{Roberts:2012aq} 
  M.~M.~Roberts,
  ``Time evolution of entanglement entropy from a pulse,''
  JHEP {\bf 1212}, 027 (2012)
  doi:10.1007/JHEP12(2012)027
  [arXiv:1204.1982 [hep-th]].
  
  
\bibitem{Mikhailov:2003er} 
  A.~Mikhailov,
  ``Nonlinear waves in AdS / CFT correspondence,''
  hep-th/0305196.
  
  
\bibitem{Gubser:2006bz} 
  S.~S.~Gubser,
  ``Drag force in AdS/CFT,''
  Phys.\ Rev.\ D {\bf 74}, 126005 (2006)
  doi:10.1103/PhysRevD.74.126005
  [hep-th/0605182].
  

\bibitem{Nayak:2018qej} 
  P.~Nayak, A.~Shukla, R.~M.~Soni, S.~P.~Trivedi and V.~Vishal,
  ``On the Dynamics of Near-Extremal Black Holes,''
  JHEP {\bf 1809}, 048 (2018)
  doi:10.1007/JHEP09(2018)048
  [arXiv:1802.09547 [hep-th]].
  

\bibitem{Poojary:2018ronp}
  R. R. Poojary,
   ``$AdS_3$ dynamics and chaos''
   to appear
   

\bibitem{Kabat:1992tb} 
  D.~N.~Kabat and M.~Ortiz,
  ``Eikonal quantum gravity and Planckian scattering,''
  Nucl.\ Phys.\ B {\bf 388}, 570 (1992)
  doi:10.1016/0550-3213(92)90627-N
  [hep-th/9203082].
  

\bibitem{Kumar:2012ui} 
  S.~P.~Kumar,
  ``Heavy quark density in N=4 SYM: from hedgehog to Lifshitz spacetimes,''
  JHEP {\bf 1208}, 155 (2012)
  doi:10.1007/JHEP08(2012)155
  [arXiv:1206.5140 [hep-th]].
    
  
\bibitem{Faedo:2014ana} 
  A.~F.~Faedo, A.~Kundu, D.~Mateos and J.~Tarrio,
  ``(Super)Yang-Mills at Finite Heavy-Quark Density,''
  JHEP {\bf 1502}, 010 (2015)
  doi:10.1007/JHEP02(2015)010
  [arXiv:1410.4466 [hep-th]].
  
  
\bibitem{deBoer:2008gu} 
  J.~de Boer, V.~E.~Hubeny, M.~Rangamani and M.~Shigemori,
  ``Brownian motion in AdS/CFT,''
  JHEP {\bf 0907}, 094 (2009)
  doi:10.1088/1126-6708/2009/07/094
  [arXiv:0812.5112 [hep-th]].
  
  
\bibitem{Son:2009vu} 
  D.~T.~Son and D.~Teaney,
  ``Thermal Noise and Stochastic Strings in AdS/CFT,''
  JHEP {\bf 0907}, 021 (2009)
  doi:10.1088/1126-6708/2009/07/021
  [arXiv:0901.2338 [hep-th]].
  
  
\bibitem{Karch:2002sh} 
  A.~Karch and E.~Katz,
  ``Adding flavor to AdS / CFT,''
  JHEP {\bf 0206}, 043 (2002)
  doi:10.1088/1126-6708/2002/06/043
  [hep-th/0205236].
  
  
\bibitem{Das:2010yw} 
  S.~R.~Das, T.~Nishioka and T.~Takayanagi,
  ``Probe Branes, Time-dependent Couplings and Thermalization in AdS/CFT,''
  JHEP {\bf 1007}, 071 (2010)
  doi:10.1007/JHEP07(2010)071
  [arXiv:1005.3348 [hep-th]].
  
\bibitem{Sekino:2008he} 
  Y.~Sekino and L.~Susskind,
  ``Fast Scramblers,''
  JHEP {\bf 0810}, 065 (2008)
  doi:10.1088/1126-6708/2008/10/065
  [arXiv:0808.2096 [hep-th]].
      
      
\bibitem{Lashkari:2011yi} 
  N.~Lashkari, D.~Stanford, M.~Hastings, T.~Osborne and P.~Hayden,
  ``Towards the Fast Scrambling Conjecture,''
  JHEP {\bf 1304}, 022 (2013)
  doi:10.1007/JHEP04(2013)022
  [arXiv:1111.6580 [hep-th]].
  
  
\bibitem{DeWolfe:2001pq} 
  O.~DeWolfe, D.~Z.~Freedman and H.~Ooguri,
  ``Holography and defect conformal field theories,''
  Phys.\ Rev.\ D {\bf 66}, 025009 (2002)
  doi:10.1103/PhysRevD.66.025009
  [hep-th/0111135].
  
  
\bibitem{Karch:2007pd} 
  A.~Karch and A.~O'Bannon,
  ``Metallic AdS/CFT,''
  JHEP {\bf 0709}, 024 (2007)
  doi:10.1088/1126-6708/2007/09/024
  [arXiv:0705.3870 [hep-th]].
  
  
\bibitem{Erdmenger:2007bn} 
  J.~Erdmenger, R.~Meyer and J.~P.~Shock,
  ``AdS/CFT with flavour in electric and magnetic Kalb-Ramond fields,''
  JHEP {\bf 0712}, 091 (2007)
  doi:10.1088/1126-6708/2007/12/091
  [arXiv:0709.1551 [hep-th]].
  
  
\bibitem{Albash:2007bq} 
  T.~Albash, V.~G.~Filev, C.~V.~Johnson and A.~Kundu,
  ``Quarks in an external electric field in finite temperature large N gauge theory,''
  JHEP {\bf 0808}, 092 (2008)
  doi:10.1088/1126-6708/2008/08/092
  [arXiv:0709.1554 [hep-th]].
  

\bibitem{Alam:2012fw} 
  M.~S.~Alam, V.~S.~Kaplunovsky and A.~Kundu,
  ``Chiral Symmetry Breaking and External Fields in the Kuperstein-Sonnenschein Model,''
  JHEP {\bf 1204}, 111 (2012)
  doi:10.1007/JHEP04(2012)111
  [arXiv:1202.3488 [hep-th]].
  
  
\bibitem{Sonner:2012if} 
  J.~Sonner and A.~G.~Green,
  ``Hawking Radiation and Non-equilibrium Quantum Critical Current Noise,''
  Phys.\ Rev.\ Lett.\  {\bf 109}, 091601 (2012)
  doi:10.1103/PhysRevLett.109.091601
  [arXiv:1203.4908 [cond-mat.str-el]].
  
  
\bibitem{Banerjee:2016qeu} 
  A.~Banerjee, A.~Kundu and S.~Kundu,
  ``Emergent Horizons and Causal Structures in Holography,''
  JHEP {\bf 1609}, 166 (2016)
  doi:10.1007/JHEP09(2016)166
  [arXiv:1605.07368 [hep-th]].
  
  
\bibitem{Perlmutter:2016pkf} 
  E.~Perlmutter,
  ``Bounding the Space of Holographic CFTs with Chaos,''
  JHEP {\bf 1610}, 069 (2016)
  doi:10.1007/JHEP10(2016)069
  [arXiv:1602.08272 [hep-th]].
  
\bibitem{Kundu:2015qda} 
  A.~Kundu,
  ``Effective Temperature in Steady-state Dynamics from Holography,''
  JHEP {\bf 1509}, 042 (2015)
  doi:10.1007/JHEP09(2015)042
  [arXiv:1507.00818 [hep-th]].
  

\bibitem{Roberts:2014ifa} 
  D.~A.~Roberts and D.~Stanford,
  ``Two-dimensional conformal field theory and the butterfly effect,''
  Phys.\ Rev.\ Lett.\  {\bf 115}, no. 13, 131603 (2015)
  doi:10.1103/PhysRevLett.115.131603
  [arXiv:1412.5123 [hep-th]].
  
  
\bibitem{Fitzpatrick:2016thx} 
  A.~L.~Fitzpatrick and J.~Kaplan,
  ``A Quantum Correction To Chaos,''
  JHEP {\bf 1605}, 070 (2016)
  doi:10.1007/JHEP05(2016)070
  [arXiv:1601.06164 [hep-th]].
  
  

\bibitem{Bhattacharya:2018nrw} 
  R.~Bhattacharya, D.~P.~Jatkar and A.~Kundu,
  ``Chaotic Correlation Functions with Complex Fermions,''
  arXiv:1810.13217 [hep-th].
  
  
\bibitem{Mandal:2017thl} 
  G.~Mandal, P.~Nayak and S.~R.~Wadia,
  ``Coadjoint orbit action of Virasoro group and two-dimensional quantum gravity dual to SYK/tensor models,''
  JHEP {\bf 1711}, 046 (2017)
  doi:10.1007/JHEP11(2017)046
  [arXiv:1702.04266 [hep-th]].
  
  
\bibitem{Gaikwad:2018dfc} 
  A.~Gaikwad, L.~K.~Joshi, G.~Mandal and S.~R.~Wadia,
  ``Holographic dual to charged SYK from 3D Gravity and Chern-Simons,''
  arXiv:1802.07746 [hep-th].
  

\bibitem{Sarosi:2017ykf} 
  G.~S\'{a}rosi,
  ``AdS$_{2}$ holography and the SYK model,''
  PoS Modave {\bf 2017}, 001 (2018)
  doi:10.22323/1.323.0001
  [arXiv:1711.08482 [hep-th]].
  

  
     
\end{thebibliography}
\end{document}